\documentclass[a4paper,fleqn,usenatbib]{mnras}

\usepackage[T1]{fontenc}
\usepackage{ae,aecompl}

\usepackage{graphicx}	
\usepackage{amsmath}	
\usepackage{amssymb}	
\usepackage[utf8]{inputenc}
\usepackage{epsfig}
\usepackage{url}
\usepackage{capt-of}
\usepackage{multicol}
\usepackage{epstopdf}
\usepackage{caption}
\usepackage{textcomp}
\usepackage[usenames,dvipsnames]{color}
\usepackage{float,lscape}
\usepackage{pdfpages}
\usepackage{pdflscape}
\usepackage{afterpage}
\usepackage{rotating}
\usepackage{arydshln}

\title[KGES: angular momentum of star-forming galaxies]{The KMOS Galaxy Evolution Survey (KGES): the angular momentum of star-forming galaxies over the last $\approx$10 Gyr}

\author[Tiley et al.]{Alfred L.\ Tiley,$^{1,2}$ S. Gillman,$^{1,3,4}$ L. Cortese,$^{2,5}$  A. M. Swinbank,$^{1}$ U. {Dudzevi{\v{c}}i{\={u}}t{\.{e}}},$^{1}$ C. M. Harrison,$^{6}$ 
\newauthor I. Smail,$^{1}$ D. Obreschkow,$^{2}$ S. M. Croom,$^{7,5}$ R. M. Sharples,$^{1,8}$ A. Puglisi,$^{1}$ 
\\
$^{1}$Centre for Extragalactic Astronomy, Department of Physics, Durham University, South Road, Durham, DH1 3LE, U.K.\\
$^{2}$International Centre for Radio Astronomy Research, University of Western Australia, 35 Stirling Highway, Crawley, WA 6009, Australia\\
$^{3}$Cosmic Dawn Center (DAWN), Copenhagen, Denmark\\
$^{4}$DTU-Space, Technical University of Denmark, Elektrovej 327, DK-2800 Kgs. Lyngby, Denmark\\ 
$^{5}$ARC Centre of Excellence for All Sky Astrophysics in 3 Dimensions (ASTRO 3D), Australia\\
$^{6}$School of Mathematics, Statistics and Physics, Newcastle University, Newcastle upon Tyne NE1 7RU, UK\\
$^{7}$Sydney Institute for Astronomy (SIfA), School of Physics, University of Sydney, NSW 2006, Australia\\
$^{8}$Centre for Advanced Instrumentation, Durham University, South Road, Durham DH1 3LE UK\\
}

\date{Accepted XXX. Received YYY; in original form ZZZ}

\pubyear{2020}

\begin{document}
\label{firstpage}
\pagerange{\pageref{firstpage}--\pageref{lastpage}}
\maketitle

\begin{abstract}
We present the KMOS Galaxy Evolution Survey (KGES), a $K$-band Multi-Object Spectrograph (KMOS) study of the H$\alpha$ and [N {\sc ii}] emission from 288 $K$ band-selected galaxies at $1.2 \lesssim z \lesssim 1.8$, with stellar masses in the range $\log_{10}(M_{*}/\rm{M}_{\odot})\approx9$\,--\,$11.5$. In this paper, we describe the survey design, present the sample, and discuss the key properties of the KGES galaxies. We combine KGES with appropriately matched samples at lower redshifts from the KMOS Redshift One Spectroscopic Survey (KROSS) and the SAMI Galaxy Survey. Accounting for the effects of sample selection, data quality, and analysis techniques between surveys, we examine the kinematic characteristics and angular momentum content of star-forming galaxies at $z\approx1.5$, $\approx1$ and $\approx0$. We find that stellar mass, rather than redshift, most strongly correlates with the disc fraction amongst star-forming galaxies at $z \lesssim 1.5$, observing only a modest increase in the prevalence of discs between $z\approx1.5$ and $z\approx0.04$ at fixed stellar mass. Furthermore, typical star-forming galaxies follow the same median relation between specific angular momentum and stellar mass, regardless of their redshift, with the normalisation of the relation depending more strongly on how disc-like a galaxy's kinematics are. This suggests that massive star-forming discs form in a very similar manner across the $\approx$ 10 Gyr encompassed by our study and that the inferred link between the angular momentum of galaxies and their haloes does not change significantly across the stellar mass and redshift ranges probed in this work.
\end{abstract}

\begin{keywords}
galaxies: general, galaxies: evolution, galaxies: kinematics and dynamics, galaxies: star formation 
\end{keywords}



\section{Introduction}
\label{sec:intro}

Galaxies are thought to have first formed as gas fell into haloes in the early Universe, mixing with the enclosed dark matter. The gas subsequently cooled, decoupling from the dark matter and collapsing to form stars and, eventually, galaxies \citep[e.g.][]{Fall:1980}. Whether or not (and how quickly) a disc forms during this process, and for how long it survives, depends on the initial angular momentum of the baryons and its distribution, as well as any subsequent redistribution during the lifetime of the galaxy \citep[e.g.][]{Freeman:1970,Fall:1983}. The former should be linked to the angular momentum of the host halo \citep[acquired via tidal torques;][]{Peebles:1969} and the initial collapse of the gas \citep[e.g.][]{Mo:1998}, whilst the latter may occur afterwards through a number of key physical processes such as inflows \citep[e.g.][]{Pichon:2011,Codis:2012,Laigle:2015}, outflows \citep[e.g.][]{Maller:2002,Dutton:2009,Dutton:2011b,Brook:2011}, or merging events \citep[e.g.][]{Welker:2017,Grand:2017}. Thus the angular momentum content of a galaxy, and whether or not it exhibits a disc structure, should be intimately connected to its initial formation {\it and} its subsequent assembly history. To further our understanding of galaxy formation and growth, it is therefore vital to be able to measure and understand the evolution of the angular momentum of galaxies over cosmic history, as well as to what extent the prevalence of discs in galaxies changes over the same period. 

Previous studies have examined the angular momentum content of galaxies in the local Universe, revealing strong correlations between the stellar specific (i.e. per unit mass) angular momentum ($j_{*}$) of a galaxy and its total stellar mass ($M_{*}$), as well as its morphology at fixed stellar mass. For example \citet{Romanowsky:2012} and \citet{Fall:2013} show that, for $\approx100$ galaxies at $z\approx0$, the normalisation of the $j_{*}$--$M_{*}$ relation is a function of galaxies' Hubble T-type and bulge-to-total ratio \citep[see also][]{Bertola:1975,Fall:1983}; disc-dominated galaxies and bulge-dominated galaxies follow parallel, but offset, versions of the $j_{*}$--$M_{*}$ relation. This aligns with the expectation that the specific angular momentum of a galaxy is linked to its formation history. Spheroidal early-type galaxies are traditionally thought to have undergone a series of major and/or minor dispersive merging events that reduce their net angular momentum \citep[e.g][]{Meza:2003,Hopkins:2010} in comparison to disc-dominated late-type galaxies \citep[although secular redistribution of angular momentum via disc instabilities and clump formation could also play a role in producing spheroids and bulges in galaxies; e.g.][]{vandenBosch:1998,Immeli:2004,Bournaud:2014}.

The maturation of optical integral field spectroscopy (IFS) technology, and its combination with multiplexed observations in the past decades, has allowed for the efficient mapping of the spatially-resolved kinematics of large numbers of galaxies using their nebular line emission and stellar absorption lines. Large optical IFS surveys such as the ATLAS$^{3\rm{D}}$ project \citep{Cappellari:2011}, the Calar Alto Legacy Integral Field Area \citep[CALIFA;][]{Sanchez:2012} survey, the Sydney-Australian-Astronomical Observatory Multi-object Integral-field Spectrograph \citep[SAMI;][]{Croom:2012} Galaxy Survey \citep{Bryant:2015}, and the Mapping Nearby Galaxies at Apache Point Observatory \citep[MaNGA;][]{Bundy:2015} survey have now observed thousands of nearby galaxies. These observations provide statistically large and well-selected data sets for the detailed study of spatially-resolved kinematics of galaxies at $z\approx0$ that span the whole range of galaxy morphologies. These have been used to great effect to extend our understanding of the link between angular momentum and the formation histories of nearby galaxies \citep[e.g.][]{Emsellem:2011,Krajnovic:2013,Cortese:2016,Greene:2018,Graham:2018,FalconBarroso:2019}. 

Similar advances in {\it near-infrared} (NIR) IFS technology have recently opened a parallel window on the kinematics of galaxies at $z\approx1$--$3$, where the well-understood rest-frame optical nebular emission lines of star-forming galaxies are redshifted into the NIR. Large IFS surveys with the European Southern Observatory's (ESO) $K$-band Multi-Object Spectrograph \citep[KMOS;][]{Sharples:2013} including the KMOS Redshift One Spectroscopic Survey \citep[KROSS;][]{Stott:2016,Harrison:2017}, the KMOS$^{3\rm{D}}$ survey \citep{Wisnioski:2015,Wisnioski:2019}, and the KMOS Deep Survey \citep[KDS;][]{Turner:2017}, in addition to surveys with other similar instruments such as the Spectroscopic Imaging survey in the near-infrared with SINFONI \citep[SINS;][]{ForsterSchreiber:2009}, have together now mapped the gas kinematics of thousands of star-forming galaxies between $z\approx0.9$ and $\approx3$ (see \citealp{ForsterSchreiber:2020} for an extensive review on this topic). These high-redshift IFS samples provide the opportunity for a thorough examination of the angular momentum content and kinematic properties of star-forming galaxies in the last $\approx11$ Gyr, spanning the epoch of peak star-formation rate density in the Universe when the majority of the stellar mass in today's Universe was assembled. Importantly, however, they also allow for direct and statistically robust kinematic comparisons between the more distant galaxy populations and galaxies in the present day, helping to provide crucial insights into how galaxies have changed over $\approx$80 per cent of the history of the Universe. 

Thanks to the work of the NIR IFS surveys discussed above, as well other parallel techniques and analyses, it is now widely accepted that star-forming galaxies in the past assembled stars at more prolific rates than their local Universe counterparts \citep[e.g.][]{Madau:2014}, with (ionised gas) kinematics consistent with turbulent discs or irregular systems, and visual morphologies that appear increasingly ``clumpy" and irregular in optical and NIR imaging with increasing redshift \citep[e.g.][]{Driver:1995, Schade:1995, Abraham:1996, vandenBergh:1996, Masafumi:1998, Dickinson:2000, Conselice:2005b, Buitrago:2013}. However, there have been relatively few studies to date that focus specifically on the angular momentum of high-redshift galaxies \citep[e.g.][]{ForsterSchreiber:2006, Burkert:2016, Contini:2016, Gillman:2019}, and even fewer that consider statistically large samples of galaxies at distant epochs. Notable examples of the latter include \citet{Harrison:2017} and \citet{Swinbank:2017}, who examined the $j_{*}-M_{*}$ relation for samples of respectively 586 H$\alpha$-detected star-forming galaxies at $z=0.6$--$1$ from KROSS, and 405 star-forming galaxies at $z=0.28$--$1.65$ observed with KMOS or MUSE. \citet{Harrison:2017} found that $z\approx0.9$ star-forming galaxies follow a $j_{*}-M_{*}$ relation that is approximately parallel to that for $z=0$ spiral galaxies, but offset lower in its normalisation by $\approx0.2$--$0.3$ dex. Similarly, \citet{Swinbank:2017} find that the same relation evolves as $j_{*} \propto M_{*}^{2/3}(1+z)^{-1}$. Whilst these studies have taken considerable steps toward a clearer understanding of the angular momentum content and kinematic properties of galaxies beyond our local Universe, several outstanding issues still persist. 

Firstly, gaps remain in high redshift IFS coverage. In particular, galaxies at $z\approx1.5$, {\it the} peak in cosmic star-formation rate density, have redshifted H$\alpha$ emission that falls within the $H$ band, which suffers from stronger sky contamination than adjacent bands at bluer and redder wavelengths (corresponding to lower and higher redshifts, respectively). As such the number of galaxies at this epoch with corresponding IFS data is small in comparison to redshifts above (i.e\ $z\approx2$--$3$) and below (i.e.\ $z\approx0.9$) it. As well as being a key period for mass assembly, this epoch also corresponds to the point at which disc morphologies start to truly emerge, eventually dominating the star-forming population \citep[e.g.][]{Mortlock:2013}. It is thus clearly a vital period in cosmic history in which to examine galaxies' kinematic properties, but lacks the large IFS samples needed to do so. 

Secondly, we are also so far missing a truly fair and direct comparison of galaxy kinematics over a large redshift range, with statistically large samples at each epoch. Existing large IFS surveys at both high and low redshift have tended to operate in isolation, conducting independent analyses with differing methodologies, measurement definitions, and analysis techniques, with only limited, and mostly indirect, comparisons of results between epochs. Given the potential for large systematic biases introduced as a result \citep[e.g.][]{Tiley:2018}, a homogeneous and unifying approach is required for a fair comparison of galaxies between redshifts.

To address these outstanding issues, in this work we present the KMOS Galaxy Evolution Survey (KGES). KGES aims to study the spatially-resolved gas properties and kinematics of a statistically large and representative sample of ``normal" star-forming galaxies at $z\approx1.5$. KGES is a Durham University guaranteed time survey with the ESO KMOS on the Very Large Telescope, Paranal, Chile. With deep KMOS $H$ band observations, it targets the H$\alpha$ and [N {\sc ii}]6548,6583 nebular line emission from 288 massive galaxies in well-known, deep extragalactic fields. 

In this paper we describe the KGES survey design and data reduction, and we present measurements of the key properties of the KGES galaxies. We then combine the KGES sample with large and representative samples of star-forming galaxies typical for their epoch at lower redshifts, observed as part of KROSS ($z\approx0.9$) and the SAMI Galaxy Survey ($z\approx0.04$). We provide a careful and coherent {\it direct} comparison of the disc fractions and angular momentum content of normal star-forming galaxies at $z\approx1.5$, $\approx0.9$, and $\approx0.04$, matching our sample selection and analysis techniques at each redshift, and robustly accounting for differences in data quality between the three epochs. 

This paper is structured as follows: In \S~\ref{sec:data} we describe the basic design of KGES, including the target selection, observing strategy and data reduction methods. We then provide a broad overview of the KGES sample in \S~\ref{sec:sample_overview}. We present the integrated properties of KGES galaxies in \S~\ref{sec:galaxy_properties}, and their resolved properties and kinematics in \S~\ref{sec:resolved_properties}. In \S~\ref{sec:lowz_comp_and_subsamples}, we discuss the KROSS and SAMI samples and measurements, and the selection of matched sub-samples between redshifts. We present and discuss our results in \S~\ref{sec:results_and_discussion}, focussing on an examination of the positions of star-forming galaxies on the stellar specific angular momentum-stellar mass plane as a function of redshift. We provide concluding remarks in \S~\ref{sec:conclusions}. 

A Nine-Year {\it Wilkinson Microwave Anisotropy Probe} \citep[{\it WMAP}9;][]{Hinshaw:2013} cosmology is used throughout this work (Hubble constant at $z = 0$, $H_{0} = 69.3 $ km s$^{-1}$ Mpc$^{-1}$; non-relativistic matter density at $z=0$, $\Omega_{0} = 0.287$; dark energy density at $z=0$, $\Omega_{\Lambda} = 0.713$). All magnitudes are quoted in the AB system. All stellar masses assume a Chabrier \citep{Chabrier:2003} initial mass function. 

\section{Sample Selection, Observations and Data Reduction}
\label{sec:data}
\subsection{Sample Selection Criteria}
\label{subsec:targetselection}

\begin{figure}
\begin{minipage}[]{.5\textwidth}
\includegraphics[width=1.\textwidth,trim= 5 10 -10 10,clip=True]{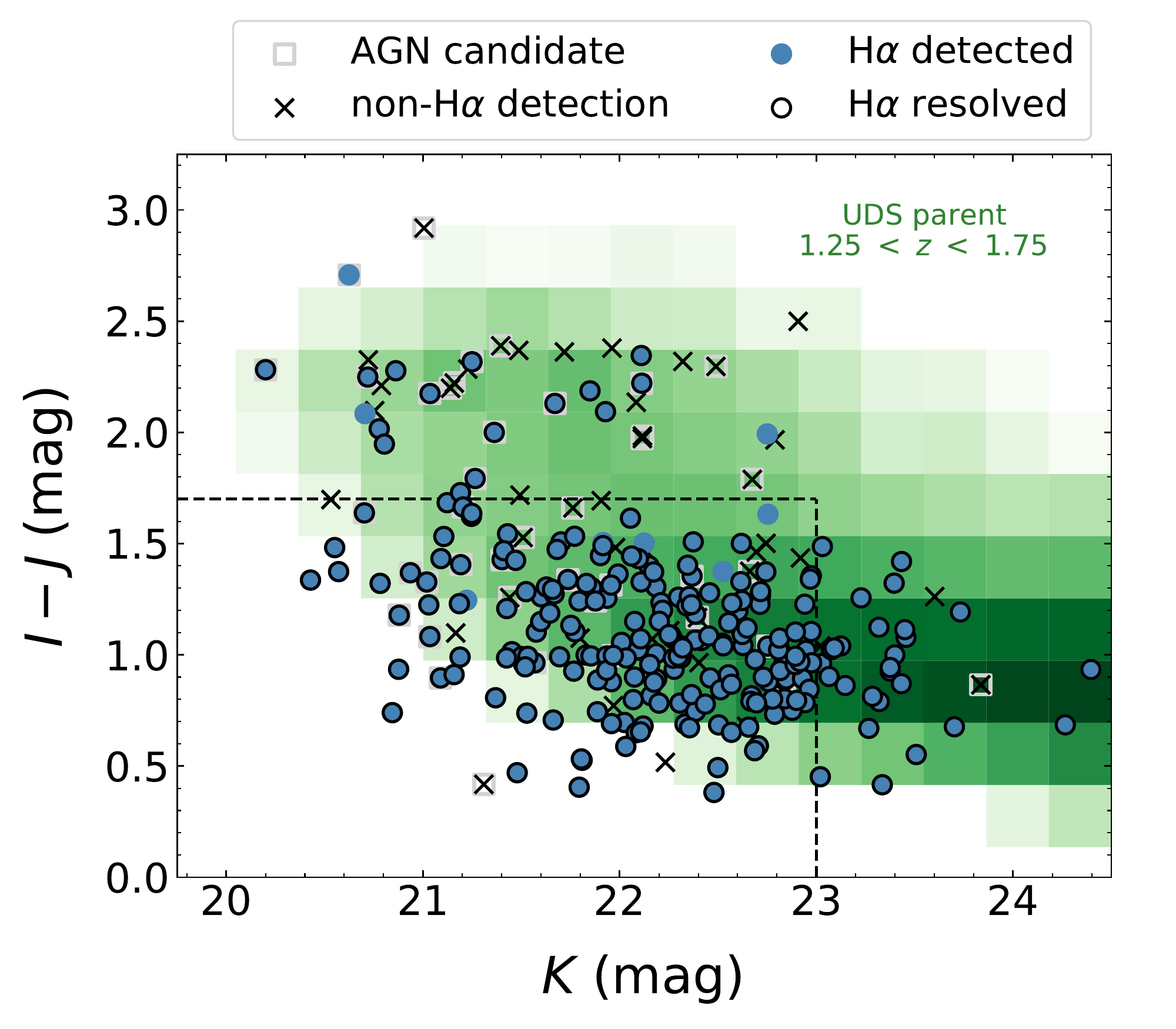}
\end{minipage}
\caption{%
The KGES galaxies in the $I-J$ versus $K$ colour-magnitude plane. H$\alpha$-detected galaxies are presented as filled blue circles. A black circular outline indicates those that are also spatially resolved in H$\alpha$ (see \S~\ref{subsec:resolved_Halpha}). Those not detected in H$\alpha$ are represented as black crosses. Candidate AGN hosts are indicated with hollow grey squares. The weak selection in magnitude ($K < 23$) is indicated by the vertical dashed line. A preference was also given to targets with $I-J < 1.7$ (horizontal dashed line). For context, we include the distribution of a ``parent" sample of $\approx24,000$ galaxies from \citet{Dudzeviciute:2020} in the UDS field (represented by a green two-dimensional histogram), with robust photometry measurements and in a similar redshift range to the KGES galaxies.  The majority of H$\alpha$-detected and resolved KGES galaxies are bright and blue. Non-detections and candidate AGN hosts tend to be redder, but span a similar range in $K$ band magnitudes as the H$\alpha$-detected galaxies. %
     }%
\label{fig:colour_magnitude}
\end{figure}

We target the H$\alpha$, [N {\sc ii}]6548 and [N {\sc ii}]6583 nebular line emission from 288 galaxies at $1.22 \leq z \leq 1.76$ in the Cosmic Evolution Survey \citep[COSMOS;][]{Scoville:2007}, Extended {\it Chandra} Deep Field South \citep[ECDFS;][]{Giacconi:2001}, and United Kingdom Infrared Telescope Deep Sky Survey \citep[UKIDSS;][]{Lawrence:2007} Ultra-Deep Survey (UDS; \citealt{Cirasuolo:2007}) fields. Of these 288, 162 (56 per cent) also fall within {\it Hubble Space Telescope} ({\it HST}) Cosmic Assembly Near-infrared Deep Extragalactic Legacy Survey \citep[CANDELS;][]{Koekemoer:2011} fields in these regions. KGES targets were preferentially selected to be bright ($K < 23$) and blue ($I-J < 1.7$) with priority given to those previously detected in H$\alpha$ emission and/or with an existing spectroscopic redshift (from MMT/Magellan Infrared Spectrograph, Hi-Z Emission Line Survey, or 3D-HST observations; \citealt{Geach:2008}; \citealt{vandokkum:2011}; \citealt{McLeod:2012}; \citealt{Chilingarian:2015}). Redder and fainter galaxies, and those without spectroscopic redshifts, were also included as lower priority targets. The distribution of KGES targets in the $I-J$ versus $K$ colour-magnitude plane is shown in Figure~\ref{fig:colour_magnitude}. 

\subsection{KMOS Observations}
\label{subsec:kmos_obs}

The KGES targets were observed over 27 nights, as part of a Durham University ESO guaranteed time observing programme. Observations were carried out with KMOS in Visitor Mode at ESO Paranal, spanning ESO observing periods P95--P100\footnote{ESO Programme IDs: 095.A-0748, 096.A-0200, 097.A-0182, 098.A-0311, and 0100.A-0134.}. Galaxies were targeted in the KMOS $H$ band to allow for detection of the galaxies' redshifted H$\alpha$ and [N {\sc ii}] emission. Targets were split across 21 unique KMOS pointings comprising 7 in ECDFS, 11 in COSMOS, and 3 in UDS. For each pointing, at least one KMOS arm was allocated to a star in order to monitor the point spread function (PSF) of the observations, and to provide a means to accurately center individual frames between observations during the data reduction. An additional number of KMOS arms in each pointing were dedicated to a complimentary observing programme, the KMOS AGN Survey at High-redshift (KASHz, \citealt{Harrison:2016}; Harrison et al., in preparation). The remaining 6--19 KMOS arms (depending on the pointing; 14 on median average) were each allocated to KGES targets. 

Observations were carried out in an ``OSOOSOOS" nod-to-sky observing pattern, where ``O" and ``S" are object (i.e. science) and sky frames, respectively -- each lasting 600 s. The total on-source exposure time for each pointing ranged from $5.4$--$32.1$ ks, with a mean of $14.8$ ks. Some targets were observed in more than one pointing and thus the total on-source exposure time for individual KGES galaxies ranges from $5.4$--$47.1$ ks and with a mean of $15.7$ ks. The final (i.e. stacked) full-width-half-maximum (FWHM) of the PSF ranged between $0\farcs5$--$0\farcs9$ in the $H$ band for individual galaxies, with a mean of $0\farcs7$. 

\subsection{KMOS Data Reduction}
\label{subsec:data_reduction}

Data cubes were reconstructed for each individual KMOS frame (i.e. each O and S frame) using the ESO {\sc esorex}\footnote{\url{http://www.eso.org/sci/software/cpl/download.html}} data reduction pipeline. The pipeline performs standard dark, flat and arc calibrations during the reconstruction, producing a $0\farcs2$ spaxel data cube for each frame. Following reconstruction we applied sky subtraction on a frame-by-frame basis, first using {\sc esorex} to perform a simple O$-$S subtraction, and then employing the Zurich Atmospheric Purge tool \citep[{\sc zap};][]{Soto:2016}, adapted for use with KMOS (Mendel et al., in preparation), to each O$-$S cube to remove residual sky contamination remaining from over- or under-subtraction in the first step. The {\sc zap} tool uses a principle component analysis to characterise and then remove the residual sky signal. 

Flux calibrations for the frames were performed using corresponding observations of standard stars taken at the same time as the target observations. And calibrated frames were centred according to the position(s) of the corresponding reference star(s) observed in each science frame. To produce final stacked cubes for galaxies observed across {\it multiple} KMOS pointings, we applied additional centering corrections according to the relative offsets between the spatial position of the peak of the galaxy's continuum emission (\S~\ref{subsec:kmosmaps}) in the stack for each of the pointings. Each galaxy is only considered once in our analysis. If a galaxy is observed across multiple KMOS pointings, we only consider the multi-pointing stack for that target in our analysis. In other words, we construct data cubes from the deepest possible observations for each galaxy in KGES.

Before further analysis, we centre the galaxy itself within its final cube based either on the position of its peak continuum  (via a two dimensional Gaussian fit to the median collapsed image of the cube; adopted for 215 -- $\approx 75$ per cent of -- KGES galaxies), or the position of the peak of its combined nebular and continuum galaxy emission (via a two dimensional Gaussian fit to a channel map extracted from the cube and centred around the nebular emission; adopted for 63 -- $\approx 22$ per cent of -- KGES galaxies). We centre a small minority of KGES galaxies (10 galaxies; $\approx 3$ per cent) within their cubes via visual inspection of both the median collapsed cube image and the nebular line emission channel map. The appropriate centering method is decided in each case after inspection by-eye of the best Gaussian fit to the median image and channel map. 

\section{Sample Overview}
\label{sec:sample_overview}



\subsection{Integrated fluxes and spectroscopic redshifts}
\label{subsec:fluxes_and_redshifts}

We measure the nebular line fluxes for each galaxy from integrated spectra, extracted from its data cube within two circular apertures with diameters of $\rm{D} = 1\farcs2$ and $\rm{D} = 2\farcs4$, respectively. We use two different aperture sizes to account for differences in the angular size and spatial distribution of the nebular flux between galaxies, finding the best compromise between maximising the signal-to-noise ratio (S/N) of the line emission and capturing as much of the galaxy's total incident flux as possible.

Before measuring the H$\alpha$ and [N {\sc ii}] flux, we first remove any detected stellar continuum emission from each spectrum by fitting and subtracting a 2$^{\rm{nd}}$ order polynomial fit, excluding the region containing the nebular emission during the fitting process. To account for the possibility of a non-perfect baseline subtraction in the region of the H$\alpha$ and [N {\sc ii}] lines, we then additionally subtract from the resultant spectrum its median value calculated in regions either side of the line emission (but not including the line region itself). 

We fit the H$\alpha$ and [N {\sc ii}]6548,6583 doublet lines in the baseline-subtracted spectrum simultaneously using a Gaussian triplet model and {\sc mpfit}\footnote{{\sc mpfit} employs $\chi^{2}$ minimisation via the Levenberg–Marquardt least-squares fitting algorithm to find the best fit model parameters.} in {\sc Python}. The three Gaussians are forced to share a common width and redshift, whilst the values of these two parameters are themselves free to vary. The intensity of the H$\alpha$ and [N {\sc ii}] doublet model components are free parameters, but the flux ratio between the two [N {\sc ii}] lines {\it within} the doublet itself is fixed so the intensity of the redder line is 2.95 times that of the bluer \citep{Acker:1989}. 

We classify a galaxy as detected in H$\alpha$ emission if its signal-to-noise $\rm{S/N}_{\rm{H}\alpha} \geq 5$ in at least one of the integrated spectra extracted from the two aperture sizes. Following the method of \citet{Stott:2016}, we calculate the signal-to-noise as 

\begin{equation}
\rm{S/N}_{\rm{H}\alpha} =  \sqrt{\chi^{2}_{\rm{base}} - \chi^{2}_{\rm{H}\alpha}}\,\,,
\end{equation}

\noindent where $\chi^{2}_{\rm{H}\alpha}$ is the chi-squared of the H$\alpha$ component of the best fit Gaussian triplet model, and $\chi^{2}_{\rm{base}}$ is the chi-squared of a horizontal line with an amplitude equal to that of the median of the baseline-subtracted spectrum in a region near to the line emission, but excluding the emission region itself \citep[e.g.][]{Neyman:1933,Bollen:1989,Labatie:2012}.

We take the spectroscopic redshift of the galaxy from the best fit to the $\rm{D} = 1\farcs2$ integrated spectrum (i.e. from the aperture that maximises $\rm{S/N}_{\rm{H}\alpha}$). For the total H$\alpha$ flux of each galaxy, we adopt the value measured from the larger, $\rm{D} = 2\farcs4$ integrated spectrum, provided we detect H$\alpha$ ($\approx 81$ per cent of targets). If H$\alpha$ emission is not detected in the $\rm{D} = 2\farcs4$ aperture but is detected in the $\rm{D} = 1\farcs2$ aperture ($\approx 3$ per cent of targets), we adopt the line flux measured from the latter but apply a multiplicative correction factor (of 1.74) calculated as the average ratio of the H$\alpha$ flux measured from the larger to the smaller aperture for those galaxies H$\alpha$-detected in both. We adopt neither measurement of H$\alpha$ flux for those galaxies with no detection in either aperture ($\approx 16$ per cent of targets).

\subsection{Detection Statistics}
\label{subsec:detection_stats}

In total, KGES targeted 288 unique galaxies with KMOS across the ECDFS, COSMOS and UDS fields. We detect H$\alpha$ emission ($\rm{S/N}_{\rm{H}\alpha} \geq 5$; \S~\ref{subsec:fluxes_and_redshifts}) in the integrated spectrum of 243 ($\approx$ 84 per cent) of these. 

\begin{figure}
\centering
\includegraphics[width=.5\textwidth]{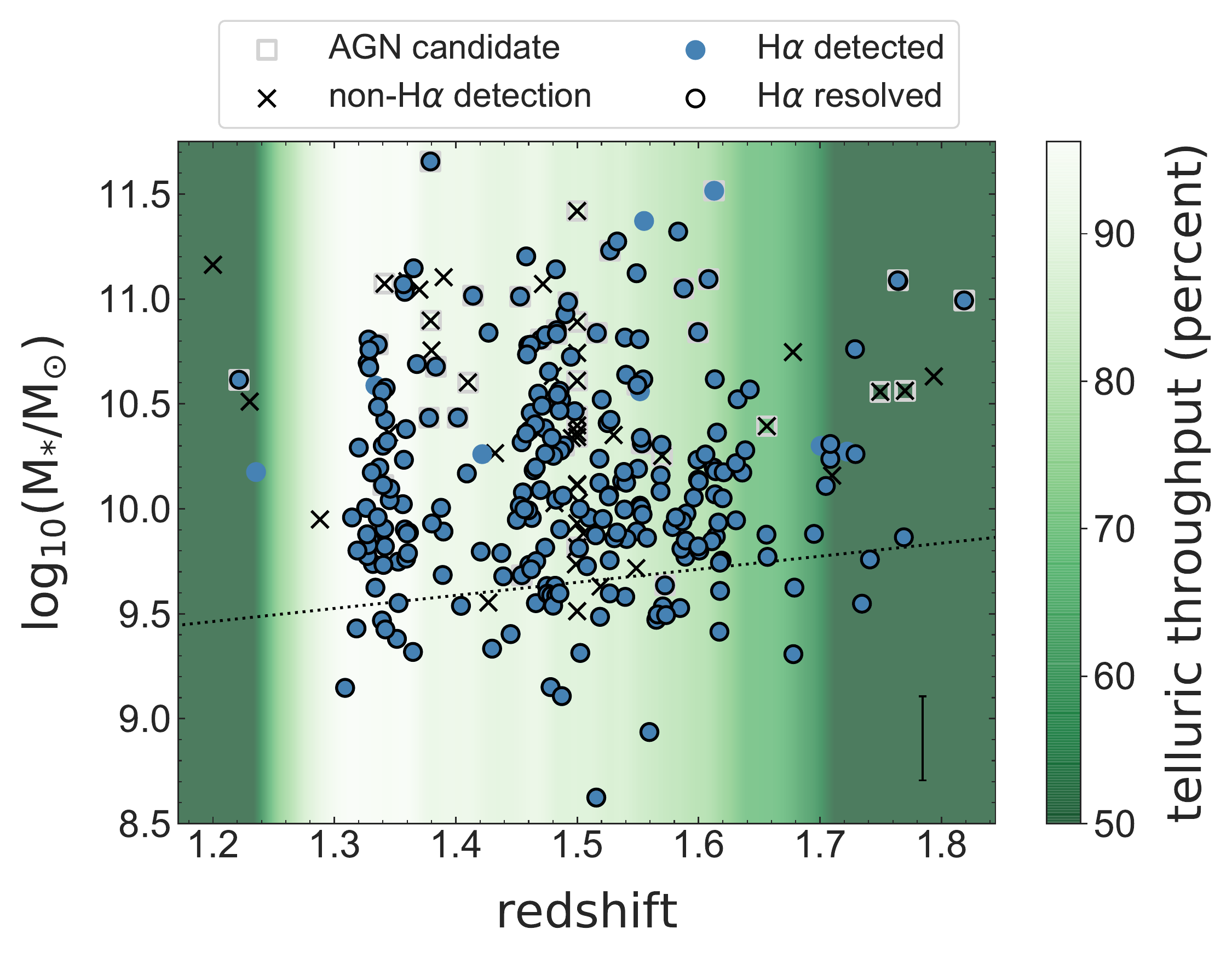}
\caption{%
The stellar masses of the KGES galaxies as a function of their redshift (spectroscopic for H$\alpha$-detected systems, photometric otherwise). Symbols are as for Figure~\ref{fig:colour_magnitude}. The stellar mass error bar (assumed constant across the sample) is shown in the bottom right. The dotted black line represents the approximate stellar mass limit for star-forming galaxies corresponding to $K = 23$ at each redshift. The green shading in the background indicates the telluric throughput at the observed wavelength of H$\alpha$ emission at the corresponding redshift (dark green to white: more severe to less sever telluric absorption). By design, the majority of H$\alpha$-detected galaxies are at redshifts corresponding to higher telluric throughput at the observed wavelength of H$\alpha$.  %
     }%
\label{fig:mass_redshift}
\end{figure}

Assuming the H$\alpha$ detections and non-detections have similar redshift distributions (see Fig.~\ref{fig:mass_redshift}), the latter are not intrinsically dimmer than the former, with the median $K$ band magnitude (and corresponding bootstrapped $1\sigma$ uncertainty) for each being $22.20 \pm 0.06$ and  $22.0 \pm 0.1$, respectively. However, the median $I-J$ colour of H$\alpha$-detected KGES galaxies is significantly bluer than non-H$\alpha$-detected systems ($1.07 \pm 0.03$ versus $1.7 \pm 0.2$, respectively, see Fig.~\ref{fig:colour_magnitude}). Thus a likely explanation for our H$\alpha$ non-detections is that these redder systems have intrinsically lower star-formation rates (which should correspond with observed colour), and thus also lower H$\alpha$ luminosities and H$\alpha$ fluxes. These systems probably fall below the H$\alpha$ flux detection limit for KGES. Alternatively they may be highly dust obscured, similarly resulting in a non-detection in H$\alpha$ (and a redder colour). 

\subsection{Identifying Candidate AGN Hosts}
\label{subsec:identifying_AGN}

For KGES, we are interested in ``normal" star-forming systems at $z\approx1.5$. Before undertaking any detailed analysis, we therefore first must ensure that the H$\alpha$ emission we detect from each KGES galaxy is driven by the photoionisation of gas surrounding young, massive stars, i.e.\ by recent or ongoing star-formation, rather than the presence of an AGN. While for some AGN reliable SFR and kinematic properties can be estimated, for this paper we decided to adopt a conservative approach and focus on bone-fide star-forming galaxies only.

The wavelength range of the KMOS $H$ band does not encompass the redshifted positions of the [O {\sc iii}] and H$\beta$ emission lines required to place the KGES galaxies on the ``Baldwin-Phillips and Terlevich" (BPT) diagram \citep{Baldwin:1981}, commonly used to indicate the presence of an AGN. We instead take an alternative, conservative approach, identifying candidate AGN hosts amongst the KGES galaxies by examining their integrated H$\alpha$ linewidths and [N {\sc ii}]/H$\alpha$ flux ratios. We also make use of ancillary {\it Spitzer}\footnote{The {\it Spitzer} IRAC/MUSYC Public Legacy Survey in the Extended CDF-South \citep[SIMPLE;][]{Damen:2011} photometry catalog, the COSMOS {\it Spitzer} survey \citep[S-COSMOS;][]{Sanders:2007} IRAC Photometry Catalog, and the {\it Spitzer} UKIDSS Ultra Deep Survey \citep[SpUDS;][]{Dunlop:2007} IRAC Catalog.} and {\it WISE}\footnote{The All{\it WISE} Source Catalog \citep{Cutri:2013a,Cutri:2013b}.} near-infrared data available for KGES galaxies, and various X-ray catalogues\footnote{The 2 Ms Point-source Catalogs for ECDFS \citep{Luo:2008}, the {\it Chandra}-COSMOS Legacy Survey Point Source Catalog \citep{Civano:2016}, and the {\it Chandra} Legacy Survey of the UKIDSS Ultra Deep Survey Field \citep[X-UDS;][]{Kocevski:2018} catalog.}. Using these sources we identify the following candidate AGN hosts in the KGES sample:

\begin{itemize}
\item 15 galaxies with a corresponding X-ray source within $1\farcs5$ with a luminosity $L_{X} \geq 10^{42}$ erg s$^{-1}$.
\item 4 galaxies with {\it Spitzer} $[5.8]-[3.6]$ and $[8.0]-[4.5]$ colours indicative of the presence of an AGN, according to the widely adopted \citet{Donley:2012} {\it Spitzer} colour selection criteria for AGN. 
\item 25 galaxies with a {\it WISE} $W1$ and $W2$ band colour (corresponding to [3.6]-[4.5]), $W1 - W2 > 0.8$ \citep{Stern:2012}. 
\item 2 galaxies with [N {\sc ii}]/H$\alpha > 0.8$ \citep[e.g.][]{Wisnioski:2018} in their integrated KMOS spectrum (extracted from the $D = 1\farcs2$ circular aperture, since the influence of the AGN should be strongest in the central regions of the galaxy).
\item 1 galaxy detected in H$\alpha$ and with an integrated FWHM H$\alpha$ line width greater than 1000 km s$^{-1}$ \citep[e.g.][]{Genzel:2014}.
\end{itemize}

In total we identify 41 ($\approx 14$ per cent) unique candidate AGN hosts in the KGES sample. Only 6 of these are flagged as AGN via more than one criterion. In Table~\ref{tab:kgesvals} we provide the AGN flag for each KGES galaxy. We detect H$\alpha$ from 26 out of the 41 ($\approx 63$ per cent) candidate AGN hosts, meaning $\approx 11$ per cent of H$\alpha$-detected KGES galaxies may host an AGN. The AGN fraction in KGES is lower than the 25 per cent measured by \citet{ForsterSchreiber:2019} for ``normal" galaxies with stellar masses in the range $\log_{10}(\rm{M}_{*}/\rm{M}_{\odot}) = 9.0$--$11.7$, at $0.6 < z < 2.7$. However, their sample extends to larger stellar masses (with a larger fraction of more massive galaxies) than KGES, where one might expect to find a higher frequency of galaxies that host a bright AGN. 

Finally, we note that our AGN selection criteria are likely to be most sensitive to {\it strong} AGN activity, dominating the bulk of the H$\alpha$ emission in our galaxies. Thus, we cannot rule out the possibility of {\it weak} AGN activity contributing to the H$\alpha$ (and [N {\sc ii}]) emission that we detect from KGES galaxies (e.g., with AGN emission only in the central spaxel), including those not flagged as candidate AGN hosts. 

\section{Integrated Galaxy Properties}
\label{sec:galaxy_properties}


\subsection{Stellar Masses}
\label{subsec:stellarmasses}

The derivation of stellar masses for the KGES sample is described in detail in \citet{Gillman:2020}. In summary, a stellar mass estimate for each KGES galaxy was obtained via the application,  in \citet{Dudzeviciute:2020}, of the Multi-wavelength Analysis of Galaxy Physical Properties \citep[{\sc magphys};][]{Cunha:2008} SED fitting routine to model its SED. Each SED itself was constructed from extensive multi-wavelength photometry spanning the ultra-violet ($UV$) to the mid-infrared ($8 \mu$m). The {\sc magphys} routine compares the observed galaxy SED to a suite of model SEDs built using the \citet{Bruzual:2003aa} spectral libraries, allowing for absorption of $UV$ light by dust and its re-emission in the infrared according to the \citet{Charlot:2000} prescription for dust attenuation of starlight. It assumes a \citet{Chabrier:2003} initial mass function, and allows for a wide variety of continuous star-formation histories with additional episodes of ``bursty" stellar assembly.

The {\sc magphys}-derived stellar masses for the KGES galaxies are shown as a function of their redshifts in Figure~\ref{fig:mass_redshift}. We also show the average telluric throughput at the observed wavelength of H$\alpha$ for the corresponding redshift. The stellar masses are in the range $\log_{10}(M_{*}/\rm{M}_{\odot})=8.62$--$11.66$, with a median $\log_{10}(M_{*}/\rm{M}_{\odot})$ of $10.14 \pm 0.04$ and a scatter of $\sigma_{\rm{MAD}} \equiv 1.483 \times \rm{MAD} = 0.52 \pm 0.04$ dex, where $\rm{MAD}$ is the median absolute deviation from the median itself. The majority of targets have redshifts corresponding to high telluric throughput (by design). On average, the H$\alpha$ non-detections and candidate AGN hosts have higher stellar masses than the ``normal" star-forming (H$\alpha$-detected) KGES galaxies. 

\subsection{H$\alpha$ Luminosities and Star-formation Rates}
\label{subsec:totfluxes_and_sfrs}

\begin{figure}
\begin{minipage}[]{.5\textwidth}
\includegraphics[width=1.\textwidth,trim= 33 10 -10 0,clip=True]{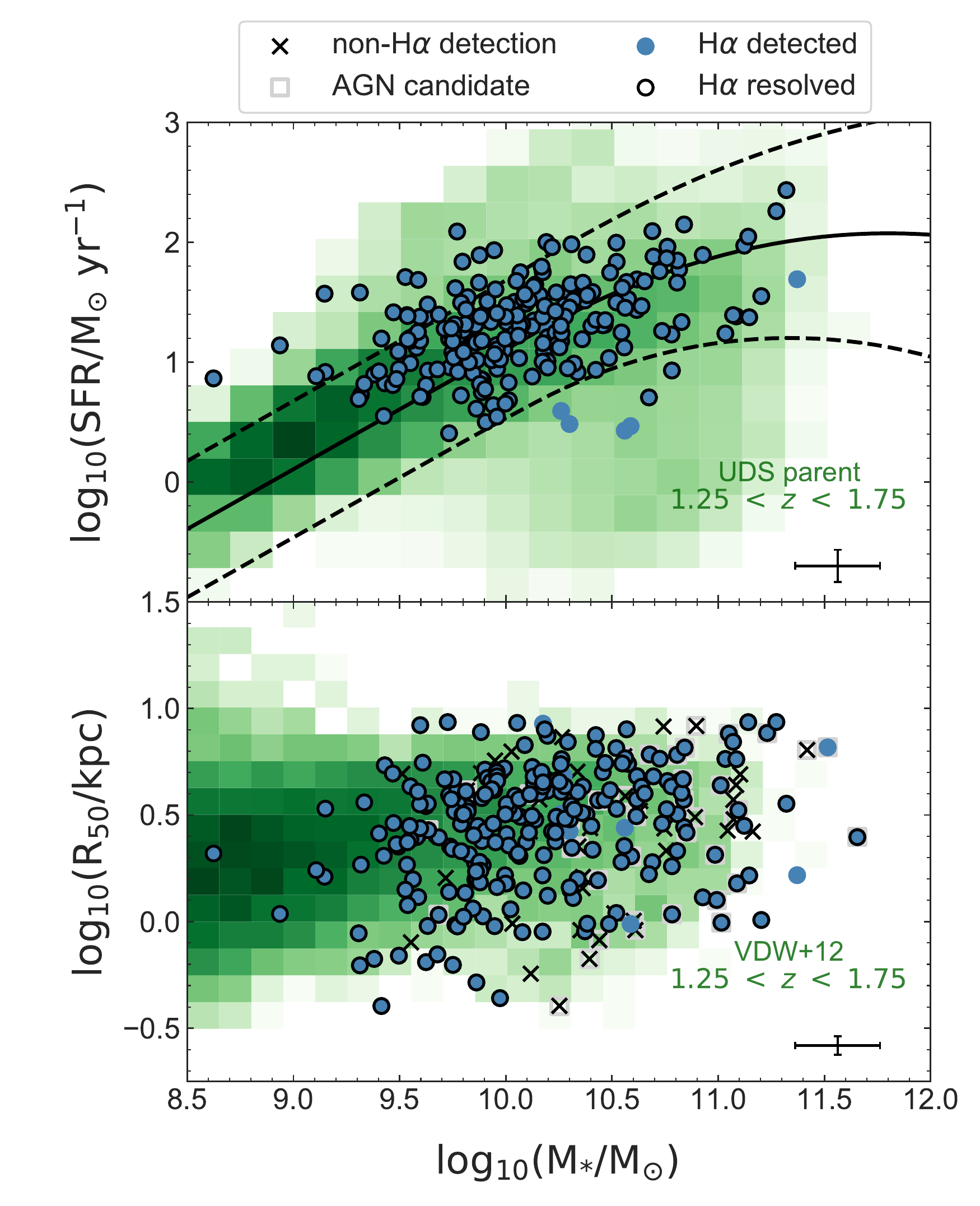}
\end{minipage}
\caption{%
{\bf Top panel:} The H$\alpha$-derived SFRs of KGES galaxies detected in H$\alpha$ emission (excluding candidate AGN hosts) as a function of their stellar masses. Symbols are as for Figure~\ref{fig:colour_magnitude}. The median error bar for the KGES points is shown in the bottom right. For context, we include a two-dimensional histogram of the positions of a ``parent" sample of galaxies in the UDS field. The  \citet{Schreiber:2015} ``main sequence" of star-formation at the median redshift of the KGES galaxies and three times its corresponding $1\sigma$ scatter are shown as respectively solid and dashed black lines. The distribution of the KGES galaxies is consistent with typical star-forming systems at the same redshift. {\bf Bottom panel:} The positions of the H$\alpha$-detected (and H$\alpha$-resolved) KGES galaxies in the stellar size-mass plane. The median error bar for the KGES points is shown in the bottom right. The spatial distribution of the KGES galaxies is in close agreement with that of a larger comparison sample of galaxies selected from CANDELS in the same redshift range ($1.25 < z < 1.75$) with sizes measured by \citet{vanderWel:2012} from $F$160$W$ ($H$ band) {\it HST} images, and stellar masses from the \citet{Santini:2015} and \citet{Nayyeri:2017} catalogues (green two dimensional histogram). The KGES galaxies have ``normal” stellar sizes for their stellar masses and redshifts. %
     }%
\label{fig:sfr_size_mass}
\end{figure}

We calculate the total H$\alpha$ luminosities and star-formation rates of the KGES galaxies based on their H$\alpha$ fluxes and redshifts. For each KGES galaxy we calculate its attenuation corrected H$\alpha$ luminosity as

\begin{equation}
L_{\rm{H}\alpha} = 4 \pi D_{L}^{2}\ 10^{0.4A_{\rm{H}\alpha,\rm{gas}}}\ F_{\rm{H}\alpha}\,\,, 
\label{eq:fHaToLHa}
\end{equation}

\noindent where $D_{L}$ is the luminosity distance calculated from the galaxy redshift, and $F_{\rm{H}\alpha}$ is the integrated H$\alpha$ flux as defined in \S~\ref{subsec:fluxes_and_redshifts}. The rest-frame nebular attenuation at the wavelength of H$\alpha$ ($A_{\rm{H}\alpha,\rm{gas}}$) is calculated according to the methods of \citet{Wuyts:2013} as 

\begin{equation}
A_{\rm{H}\alpha,\rm{gas}} = A_{\rm{H}\alpha,\rm{stars}}(1.9 - 0.15A_{\rm{H}\alpha,\rm{stars}})\,\,,
\label{eq:AstarToAgas}
\end{equation}

\noindent where $A_{\rm{H}\alpha,\rm{stars}}$ is the rest-frame stellar attenuation at the wavelength of H$\alpha$, converted from the $V$ band stellar attenuation  ($A_{V}$) assuming a \citet{Calzetti:1994} extinction law. Without H$\beta$ we cannot calculate the dust attenuation directly via the commonly used ``Balmer decrement'' (i.e.\ the H$\alpha$/H$\beta$ flux ratio). Instead we adopt the best fit $A_{V}$ from {\sc magphys}. 

The star-formation rate (SFR$_{\rm{H}\alpha}$) for each galaxy is converted from its $L_{\rm{H}\alpha}$ according to the prescription of \citet{Kennicutt:1998}, such that

\begin{equation}
\frac{\rm{SFR}_{\rm{H}\alpha}}{\rm{M}_{\odot}\ \rm{yr}^{-1}} = \rm{C}_{\rm{IMF}} \ \rm{X}_{\rm{H}\alpha} \ \frac{L_{\rm{H}\alpha}}{\rm{ergs\ s}^{-1}}\,\,,
\label{eq:LHaToSFR}
\end{equation}

\noindent where $\rm{X}_{\rm{H}\alpha} = 7.9 \times 10^{-42}\ \rm{M}_{\odot}\ \rm{yr}^{-1}\ \rm{ergs}^{-1}\ \rm{s}$ is the \citet{Kennicutt:1998} conversion factor between H$\alpha$ luminosity and star-formation rate, for a \citet{Salpeter:1955} IMF. We convert to a \citet{Chabrier:2003} IMF with a multiplicative factor of $\rm{C}_{\rm{IMF}} = 10^{-0.201}$ \citep{Madau:2014}.

The SFR$_{\rm{H}\alpha}$ for H$\alpha$-detected KGES galaxies (excluding AGN candidates) are shown as a function of their stellar masses in the upper panel of Figure~\ref{fig:sfr_size_mass}. For context, we include a two-dimensional histogram of the positions of a  
K-band selected ``parent'' sample of UDS field galaxies representative of the star-forming main sequence in a similar redshift range to the KGES galaxies ($1.25 < z < 1.75$). Star-formation rates and stellar masses are derived via {\sc magphys} as discussed in \citet{Dudzeviciute:2020}. We also include the ``main sequence" of star-formation, according to the findings of \citet{Schreiber:2015}, at the median redshift of the KGES galaxies. The distribution of the KGES galaxies is coincident with the main locus of the  \citet{Dudzeviciute:2020} comparison sample, and also coincides with the \citet{Schreiber:2015} main sequence at their median redshift, albeit with the KGES points exhibiting a slight systematic offset toward higher SFRs at fixed stellar mass in comparison to the \citet{Schreiber:2015} trend. The KGES galaxies are typical star-forming systems for their stellar masses and redshifts.

\subsection{Stellar Structural Parameters}
\label{subsec:sizes}

The stellar light structural parameters of the KGES galaxies, including their axial-ratio derived inclinations, their S\'ersic indices, and the stellar half-light radii were measured by \citet{Gillman:2020} via the application of the {\sc galfit} \citep{Peng:2010} S\'ersic modelling code to the highest resolution, deepest, and reddest-wavelength broadband image available for each galaxy. The {\sc galfit} routine accounts for the size of the image PSF in each case, providing an intrinsic best fit model of the two-dimensional stellar light distribution. 

Approximately half ($\approx 56$ per cent) of the KGES sample fall within the CANDELS footprint. The majority of these galaxies ($70$ per cent) have corresponding deep, high-resolution {\it HST} images in $F$435$W$ ($B$), $F$606$W$ ($V$), $F$814$W$ ($I$), $F$105$W$ ($Y$), $F$125$W$ ($J$), and $F$160$W$ ($H$) bands. The remainder only have corresponding $F$435$W$, $F$606$W$, $F$814$W$ imaging. An extra 6 KGES galaxies have either $F$125$W$ or $F$125$W$ archival {\it HST} imaging.

Archival {\it HST} $F$814$W$ band imaging is also available for a further third ($\approx 31$ per cent) of the sample. For those galaxies, a correction is applied to their S\'ersic indices and half-light radii based on the average ratio of the respective values measured in $F$160$W$ band imaging to those measured in the $F$814$W$ band for those galaxies imaged in both. For the remaining minority of the sample ($\approx11$ per cent), we rely on ground-based $H$ or $K$ band imaging to measure their stellar light structural properties. 

The stellar-half light radii of the KGES galaxies are shown as a function of their stellar masses in the lower panel of Figure~\ref{fig:sfr_size_mass}. The positions of the KGES galaxies in the stellar size-stellar mass plane are in good agreement with those of a larger sample of galaxies in the CANDELS fields in a similar redshift interval ($1.25 < z < 1.75$), spatially-resolved in {\it HST} imaging, with sizes measured by \citet{vanderWel:2012}, and stellar masses from the \citet{Santini:2015} and \citet{Nayyeri:2017} catalogues for respectively ECDFS and UDS, and the COSMOS field. The KGES galaxies have stellar sizes that are ``normal" for their stellar masses and redshifts.

\section{Resolved Galaxy Properties and Kinematic Measurements}
\label{sec:resolved_properties}


\subsection{Resolved KMOS Maps}
\label{subsec:kmosmaps}

To construct maps of galaxy properties from the KMOS observations we first model and subtract the nebular emission in the centred data cubes, on a spaxel-by-spaxel basis. Since the relative contributions of noise and sky contamination are higher in the spectra of individual spaxels in the cube than in the integrated spectra described in \S~\ref{subsec:fluxes_and_redshifts}, we employ an adapted baseline subtraction method that differs from the one described in that section. For each spaxel, we start by dividing its spectrum into segments of 50 spectral bins. To each segment we then apply a $2\sigma$ iterative clip to remove any residual sky signal. We then fit and subtract a $3^{\rm{rd}}$ order polynomial to the clipped spectrum. After this we calculate the median of the clipped, polynomial-subtracted spectrum for regions either side of the line emission but excluding the line region itself. We construct our continuum model for the spectrum as the sum of the best fit $3^{\rm{rd}}$ order polynomial and the subsequently calculated median value. As a final step we subtract this continuum model from the {\it original}, unaltered spectrum for the spaxel and place this subtracted version in place of the original in the cube. This process is repeated for every spaxel to create a ``baseline-subtracted" cube. Before extracting maps from the baseline-subtracted cubes, we also regrid them from the native $0\farcs2$ spaxels to $0\farcs1$ spaxels, conserving the flux in each slice during the process. 

We model the H$\alpha$ and [N {\sc ii}] emission in each spaxel of the centred, baseline-subtracted, regridded cubes adopting the same model and methods as outlined in \S~\ref{subsec:fluxes_and_redshifts} and applied to the integrated spectra. To construct the maps we employ an adaptive binning process, in line with that used in the KROSS analyses \citep[e.g.][]{Stott:2016} and to construct maps for SAMI galaxies in \citet{Tiley:2018}, whereby for each spaxel we sum the flux in an increasing number of surrounding spaxels (fitting the Gaussian triplet model in each step) until a $\rm{S/N}_{\rm{H}\alpha} \geq 5$ is achieved. For each spaxel, we start by considering the flux within a $0\farcs3 \times 0\farcs3$ spatial bin centred on the spaxel in question. If $\rm{S/N}_{\rm{H}\alpha} < 5$, we then consider a $0\farcs5 \times 0\farcs5$ bin, and finally a $0\farcs7 \times 0\farcs7$. If we still do not formally detect H$\alpha$ then we mask the considered spaxel in the final maps of the emission line properties. We repeat this full process for every spaxel in the data cube, and for each galaxy.

We construct maps of H$\alpha$ intensity ($I_{\rm{H}\alpha}$), [N {\sc ii}] intensity ($I_{\rm{N}\textsc{ii}}$), observed line-of-sight velocity ($v_{\rm{obs}}$), and observed line-of-sight velocity dispersion ($\sigma_{\rm{obs}}$) from the KMOS data cubes. To do this we consider, in each spaxel for each galaxy, respectively the integral of the H$\alpha$ component, the integral of the redder [N {\sc ii}] line component, the central position of the H$\alpha$ component, and the common (sigma) width (corrected for the instrumental broadening) of the three Gaussian components in the best fit model to the observed nebular emission. The latter two quantities are converted into units of velocity in the galaxy rest frame in each case. To remove ``bad" pixels from the maps, for example where the best model fit is adversely affected by the presence of residual sky signal, as well as non-resolved features, we apply an iterative masking process, described in \citet{Tiley:2020}. 

We construct a stellar continuum map for each galaxy from the sum of the model continuum derived for each spaxel of the original $0\farcs2$ cubes as described above. To match the spatial sampling of the emission line maps, we regrid the resultant continuum map (conserving flux) from $0\farcs2$ spaxels to $0\farcs1$ spaxels.

\begin{figure*}
\rotatebox{-90}{
\begin{minipage}[]{1.25\textwidth}
\includegraphics[width=.99\textwidth,trim= 132 40 120 52,clip=True]{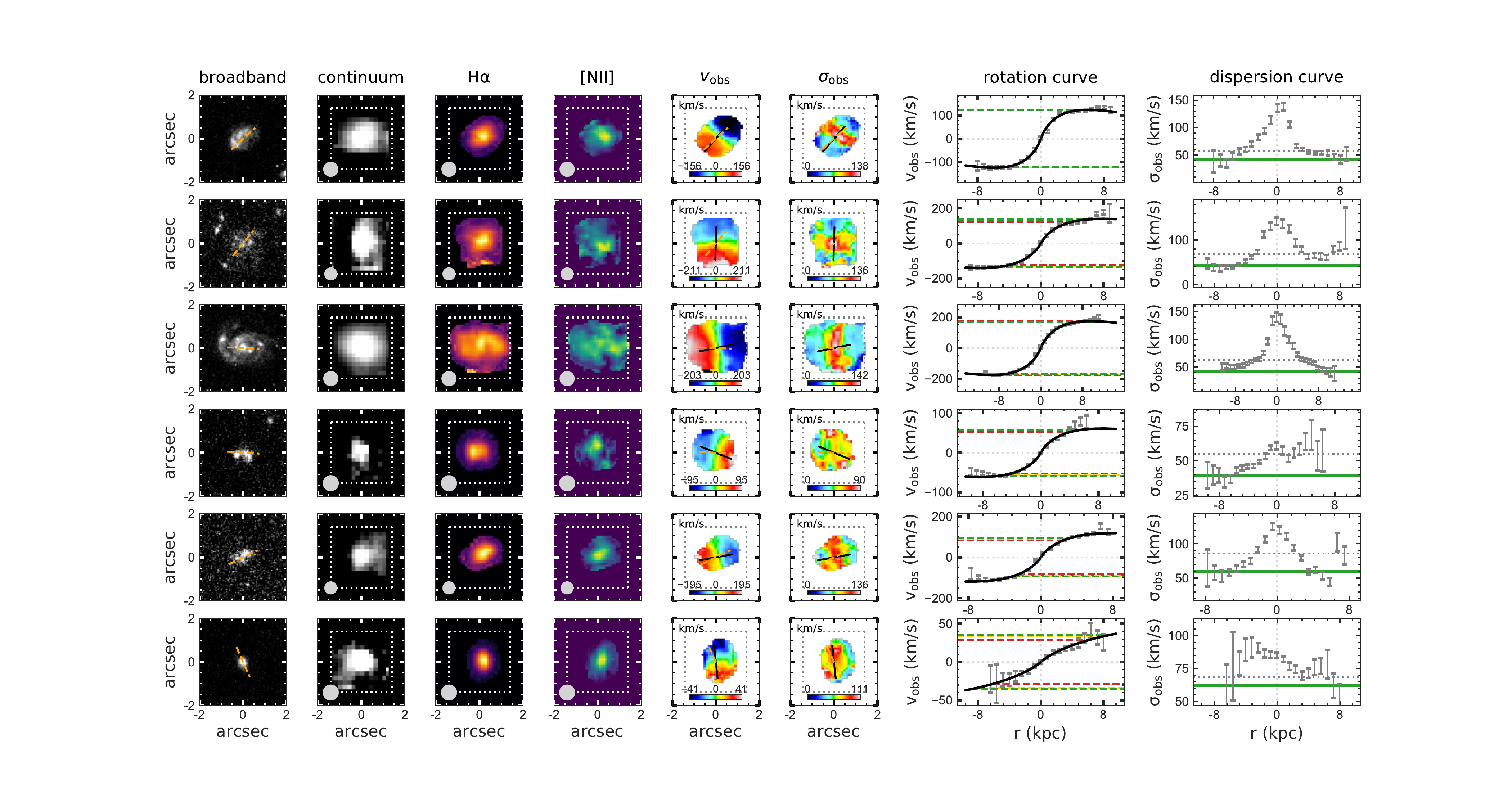}
\caption{%
Examples of spatially resolved KGES galaxies. We show one row for each galaxy with (left to right) the {\it HST} $F$814$W$ band image, KMOS continuum, H$\alpha$ intensity, [N {\sc ii}] intensity, mean line-of-sight velocity ($v_{\rm{obs}}$), and velocity dispersion ($\sigma_{\rm{obs}}$) maps, along with the observed H$\alpha$ rotation curve (extracted from the KMOS $v_{\rm{obs}}$ map along the major kinematic axis; \S~\ref{subsec:rotationvelocities}), and the observed H$\alpha$ velocity dispersion curve, extracted from the KMOS $\sigma_{\rm{obs}}$ map accordingly. The nominal KMOS FOV is indicated via a white dotted line in the KMOS continuum, H$\alpha$ and [N {\sc ii}] maps, and a grey dotted line in the KMOS $v_{\rm{obs}}$ and $\sigma_{\rm{obs}}$ maps. The size of the KMOS observation PSF for each galaxy is indicated by a grey filled circle in the bottom left of its KMOS continuum, H$\alpha$, and [N {\sc ii}] maps. The red, yellow, and green dashed lines overlaid onto the rotation curve for each galaxy represent the observed rotation velocity measured at respectively $(1.31R_{50})'$,  $(1.8R_{50})'$, and $(2R_{50})'$ (corresponding to $(2.2\rm{R}_{\rm{d}})'$, $(3\rm{R}_{\rm{d}})'$, and $(3.4\rm{R}_{\rm{d}})'$ for a pure exponential disc). For the dispersion curves, the dotted grey and solid green horizontal lines represent respectively the median value of the observed curve, and the best beam smearing-corrected measure of the intrinsic dispersion. All of the examples displayed exhibit disc-like characteristics, based on the appearance of their velocity field, rotation and dispersion curves, and their H$\alpha$ maps -- despite often appearing more ``clumpy" in the broadband {\it HST} imaging. %
     }%
\label{fig:examplemaps}
\end{minipage}}
\end{figure*}

In Figure~\ref{fig:examplemaps}, we present the constructed KMOS maps for example KGES galaxies.

\subsection{Resolved H$\alpha$ Emission}
\label{subsec:resolved_Halpha}

We classify a galaxy as spatially resolved in H$\alpha$ emission if its $I_{\rm{H}\alpha}$ map, after masking any bad pixels, contains at least one contiguous emission region with an area larger than 1.1 times the area of one resolution element (defined by the FWHM contour of the KMOS PSF). Here we adopt a 10 per cent margin of uncertainty in order to only select for those galaxies that are robustly resolved, ignoring marginal cases.  We spatially resolve the H$\alpha$ emission from 235 out of the 288 targeted KGES galaxies, corresponding to $\approx 82$ per cent of the total sample, and $\approx 97$ per cent of those from which we detect H$\alpha$. Of the 235 resolved galaxies, 25 galaxies ($\approx 11$ per cent) are flagged as candidate AGN hosts. Thus, in total, we spatially-resolve the H$\alpha$ emission in 210 ``normal" star-forming galaxies in KGES ($\approx 73$ per cent of all galaxies targeted by KGES).

\subsection{Kinematic Position Angles}
\label{subsec:PAkin}

To find the kinematic position angle for each KGES galaxy, we rotate its $v_{\rm{obs}}$ map about its centre in one degree increments. For each rotation we measure the spread of absolute velocities for pixels that fall within a horizontal, $0\farcs3$ wide ``slit" across the centre of the map (ignoring the uppermost 10$^{\rm{th}}$ per centile of values to exclude extreme outliers that may heavily bias the range). For the majority (202; $\approx 86$ per cent) of H$\alpha$-resolved KGES galaxies, we define the kinematic position angle, $\psi$, as the average of the angle of the map rotated from the horizontal that maximises the velocity spread along the slit ($\psi_{\rm{max}}$), and the angle that minimises it ($\psi_{\rm{min}}$) plus or minus 90 degrees, i.e.\  $\psi = 0.5(\psi_{\rm{max}} + \psi_{\rm{min}} \pm 90)$, see Fig.~\ref{fig:examplemaps}. In a minority of cases in which one of the two measures dramatically fails, instead of the average of the two angles, we instead adopt only one or the other; for 14 galaxies ($\approx 6$ per cent of those resolved in H$\alpha$) we adopt $\psi = \psi_{\rm{min}} \pm 90$, and for an additional 18 galaxies ($\approx 8$ per cent of resolved systems) we set $\psi = \psi_{\rm{max}}$. 

In each case the most appropriate prescription for the kinematic position angle is decided via visual inspect of the kinematic axis determined via each of the three methods overlayed onto the $v_{\rm{obs}}$ map. For a single galaxy (KGES\_284), with very complex structure in its velocity field, we manually set the kinematic position angle by-eye (to $\psi = 0$). 

\subsection{Ionised Gas Rotation Velocities}
\label{subsec:rotationvelocities}

We extract the observed rotation curve for each galaxy by rotating its $v_{\rm{obs}}$ map so that its kinematic position angle aligns with the horizontal and then calculating the weighted mean velocity (and the associated standard error) in $0\farcs1$ steps along the same $0\farcs3$-wide horizontal ``slit" as used in \S~\ref{subsec:PAkin}. Extracted rotation curves are shown for example KGES galaxies in Figure~\ref{fig:examplemaps}.

\subsubsection{Observed Rotation Velocities}
\label{subsubsec:v_obs}

To measure the observed rotation velocity of each KGES galaxy we first model its rotation curve to mitigate the effects of noise in the data. Following the methods of \citet{Harrison:2017} and \citet{Tiley:2018}, we find the best fit exponential disk model \citep[see ][]{Freeman:1970} to each rotation curve, where the model velocity as a function of galactocentric radius, $v(r)$, takes the form

\begin{equation}
(v(r) - v_{\rm{off}})^{2} = \frac{(r-r_{\rm{off}})^{2}\pi G\mu_{0}}{h}(I_{0} K_{0} - I_{1} K_{1})\,\,,
\label{eq:diskmodel}
\end{equation}

\noindent where $\mu_{0}$ and $h$ are respectively the peak mass surface density and disk scale radius, and $I_{\rm{n}}K_{\rm{n}}$ are Bessel functions evaluated at $0.5r/h$. We also include parameters to allow for a systematic offset of the rotation curve in the spatial and velocity directions; $v_{\rm{off}}$ and $r_{\rm{off}}$ are the velocity at which $r = 0$ and the radius at which $v = 0$, respectively. Each rotation curve is corrected for non-zero values of $v_{\rm{off}}$ and $r_{\rm{off}}$ before we consider it for further analysis. 

We measure the observed rotation velocities, $v_{2.2,\rm{obs}}$, for each KGES galaxy (spatially-resolved in H$\alpha$ emission) from the best fit, centred (i.e.\ $v_{\rm{off}} = 0$ and $r_{\rm{off}} = 0$) exponential disk model at $1.31R_{50}$ (corresponding to 2.2$h$ and the peak of the rotation curve for a pure disk), combined in quadrature with the sigma width of the KMOS PSF ($\sigma_{\rm{PSF}}$), i.e.\ at $(1.31R_{50})' \equiv \sqrt{(1.31R_{50})^{2} + \sigma^{2}_{\rm{PSF}}}$. 

We calculate our rotation velocities at $1.31R_{50}$ as a compromise between (1) facilitating a direct comparison between the KGES kinematics with those we measure for star-forming galaxies at $z\approx0$ in the SAMI Galaxy Survey (\citealp{Bryant:2015}, see \S~\ref{sec:lowz_comp_and_subsamples}) with H$\alpha$ kinematics typically traced out to a maximum of $\approx1.31R_{50}$, and (2) also ensuring we reach at least the turnover of the rotation curve for a measure of rotation close to that of the ``flat" outer regions. In fact, the velocity measurement is quite robust to our choice of radius; we also measure velocities at $(1.8R_{50})'$ and $(2R_{50})'$ (other commonly adopted radii in the literature) from the same best fit model curve in each case. These we label $v_{3,\rm{obs}}$ and $v_{3.4,\rm{obs}}$, respectively since they should correspond to the rotation velocity at $3h$ and $3.4h$ for a pure exponential disk. For spatially-resolved KGES galaxies we find median fractional differences of $5 \pm 1$ per cent and $6 \pm 1$ per cent between $v_{3,\rm{obs}}$ and $v_{2.2,\rm{obs}}$, and $v_{3.4,\rm{obs}}$ and $v_{2.2,\rm{obs}}$, respectively. 

\subsubsection{H$\alpha$ Extent}
\label{subsec:halpha_extent}

To understand the extent (if any) to which we must extrapolate beyond the data of each galaxy's rotation curve to measure its $v_{2.2,\rm{obs}}$, we measure the maximum radial extent of the H$\alpha$, $r_{\rm{H}\alpha,\rm{max}}$.  We define this as the maximum galactocentric radius that we detect H$\alpha$ along the major kinematic axis of each galaxy. This we simply read from each galaxy's centred (i.e\ corrected for non-zero best fit values of $v_{\rm{off}}$ and $r_{\rm{off}}$; \S~\ref{subsubsec:v_obs}) rotation curve, taking the absolute value of the maximum radial extent of the extracted curve. We do not need to extrapolate beyond the rotation curve data to measure $v_{2.2,\rm{obs}}$ (i.e.\ $r_{\rm{H}\alpha,\rm{max}}/(1.31R_{50})' \geq 0.9$) for the majority of spatially-resolved KGES galaxies ($\approx 96$ per cent). Moreover, for 45\% of our sample the rotation curve extends more than 2$\times$(1.31R$_{50}$)'.

\subsubsection{Corrected Rotation Velocities}
\label{subsubsec:v_int}

For a measure of the intrinsic rotation velocity, for each galaxy we first apply a multiplicative correction factor ($\epsilon_{\rm{R},\rm{PSF}}$) to $v_{2.2,\rm{obs}}$, according to the methods of \citet{Harrison:2017} and \citet{Johnson:2018}, to account for the effects of ``beam smearing" due to the KMOS PSF. This factor is dependent on the observed velocity shear of the galaxy and the size of the galaxy with respect to the KMOS PSF. We then apply a second, higher-order correction, based on the findings of \citet{Tiley:2018} and designed to augment the \citet{Johnson:2018} correction for improved accuracy for galaxies with intrinsically low rotation speeds or small sizes with respect to the KMOS PSF. Finally we also correct for the effects of the galaxy's inclination in each case. Our final estimate of the intrinsic rotation velocity at $1.31R_{50}$ for each KGES galaxy is thus calculated as 

\begin{equation}
v_{2.2_{\rm{C}}} = \frac{\epsilon_{\rm{R},\rm{PSF}} \times v_{2.2,\rm{obs}} + b}{m \sin\theta_{\rm{i}}}
\label{eq:V22C}
\end{equation}

\noindent where $b = 18$ km s$^{-1}$ and $m = 1.05$ \citep[][]{Tiley:2018}, and $\theta_{\rm{i}}$ is the inclination. 

The ($\log_{10}$) distribution of $v_{2.2_{\rm{C}}}$ for a subset of KGES galaxies with robust kinematics measurements (those in the {\it kinematics} sub-sample; see \S~\ref{subsec:kinematics_selection}) is shown in Figure~\ref{fig:subsample_survey_comp_histograms}. Their median $v_{2.2_{\rm{C}}}$ is 116 $\pm$ 8 km s$^{-1}$, with a scatter of $\sigma_{\rm{MAD}} = $ 64 $\pm$ 6~km~s$^{-1}$. 

\subsection{Ionised Gas Velocity Dispersions}
\label{subsec:gasdispersions}

For each resolved KGES galaxy, we extract an observed velocity dispersion curve along its major kinematic axis in the same manner as described for the rotation curve, but substituting the galaxy's $\sigma_{\rm{obs}}$ map in place of its $v_{\rm{obs}}$ map. The extracted velocity dispersion curves for example KGES galaxies are shown in Figure~\ref{fig:examplemaps}.

\subsubsection{Observed Velocity Dispersions}
\label{subsec:sigma_observed}

We define the {\it observed} velocity dispersion ($\sigma_{0,\rm{obs}}$) for each KGES galaxy in one of two ways. Either we take the median value of the points in the dispersion curve at radii $|r| > (1.31R_{50})' - 0\farcs1$, provided at least 3 points in the dispersion curve satisfy this criterion (the $0\farcs1$ buffer accounts for pixelisation of the curve). If not, or if visual inspection of the dispersion curve reveals any extremely outlying points, we instead adopt the median of the $\sigma_{\rm{obs}}$ map. We prefer the former method where possible, adopting it for $\approx 71$ per cent of resolved systems, since it is measured from pixels at larger radii that are less affected by beam smearing and thus require a smaller subsequent beam smearing correction (see \S~\ref{subsec:sigma_int}). We only adopt the median of the map for $\approx 29$ per cent of H$\alpha$-resolved KGES galaxies.

We note that, for those galaxies with sufficiently spatially-extended H$\alpha$, our measurement of the observed dispersion is robust to whether we adopt $(1.31R_{50})'$, $(1.8R_{50})'$, or $(2R_{50})'$ as our minimum radius, with a median fractional difference of $0.0 \pm 0.2$ per cent between the velocity dispersion calculated outside of either $(1.8R_{50})'$ or $(2R_{50})'$ in comparison to that calculated outside of $(1.31R_{50})'$ (with corresponding $\sigma_{\rm{MAD}}$ scatters of respectively $4 \pm 1$ per cent and $ 7 \pm 1$ per cent). We therefore adopt the smallest of the three radii to maximise the number of KGES galaxies for which we are able to make a measurement without resorting to the median of the $\sigma_{\rm{obs}}$ map.

\subsubsection{Corrected Velocity Dispersions}
\label{subsec:sigma_int}

For a characteristic measure of the {\it intrinsic} gas dispersion for each KGES galaxy, we correct the observed gas velocity dispersion (\S~\ref{subsec:sigma_observed}) for the effects of beam smearing due to the KMOS PSF. As for the observed rotation velocities, we correct the velocity dispersions in two steps. We apply a first order beam smearing correction factor ($C_{\rm{R},\rm{PSF}}$) according to the methods of \citet{Johnson:2018}, which depends on the (stellar) size of the galaxy in relation to the size of the KMOS PSF, and the velocity shear across the galaxy. We then apply a second order correction based on the findings of \citet{Tiley:2018}. The final, corrected dispersion is given as

\begin{equation}
\sigma_{0_{\rm{C}}} = \frac{C_{\rm{R},\rm{PSF}} \times \sigma_{0,\rm{obs}} + B}{M}\,\,,
\label{eq:sig0C}
\end{equation}

\smallskip

\noindent where $B = -3$ km s$^{-1}$ and $M = 1.08$ \citep{Tiley:2018}. 

\begin{figure*}
\centering
\begin{minipage}[]{1.\textwidth}
\includegraphics[width=.98\textwidth,trim= 0 0 0 0,clip=True]{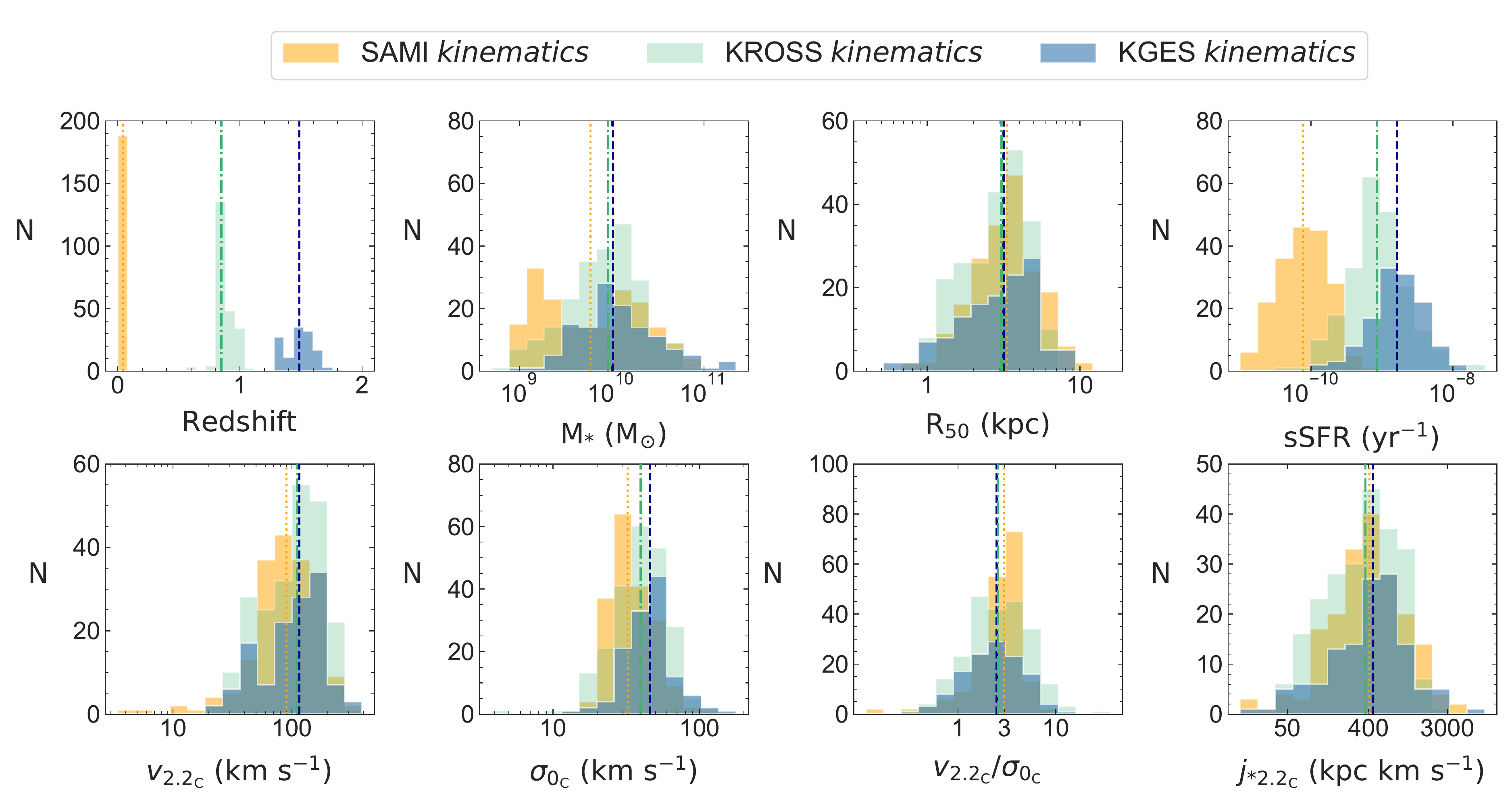}
\end{minipage}
\caption{%
The distributions of key properties of star-forming galaxies in our {\it kinematics} sub-samples at $z\approx1.5$ (KGES),  $z\approx0.9$ (KROSS), and $z\approx0.04$ (SAMI). The median of each distribution is shown as a dashed vertical line in the corresponding colour. The key properties of normal (massive) star-forming galaxies are surprisingly constant with redshift, aside from significantly elevated sSFRs and gas velocity dispersions with increasing redshift (as judged by formal comparison of the median values between redshifts, in each case). %
     }%
\label{fig:subsample_survey_comp_histograms}
\end{figure*}

The distribution of $\sigma_{0_{\rm{C}}}$ for a subset of KGES galaxies with robust kinematics measurements (the {\it kinematics} sub-sample; see \S~\ref{subsec:kinematics_selection}) is shown in Figure~\ref{fig:subsample_survey_comp_histograms}. Their median average $\sigma_{0_{\rm{C}}}$ is 46 $\pm$ 2 km s$^{-1}$, with a scatter of $\sigma_{\rm{MAD}} =$ 14~$\pm$ 1~km~s$^{-1}$. 

\subsection{Specific angular momentum}
\label{subsec:specific_angmom}

Assuming that the rotation velocity of the gas is equivalent to that of the stars, we calculate the total specific stellar angular momentum ($j_{*2.2_{\rm{C}}}$) based on the approximation devised by \citet{Romanowsky:2012}, such that 

\begin{equation}
j_{*2.2_{\rm{C}}} = k_{n} v_{2.2_{\rm{C}}} R_{50}\,\,,
\label{eq:jstar}
\end{equation}

\medskip

\noindent where $k_{n}$ is a multiplicative correction factor based on the galaxy's S\'ersic index ($n_{\rm{S}}$) and given as

\begin{equation}
k_{n} = 1.15 + 0.029n_{\rm{S}} + 0.062n^{2}_{\rm{S}}\,\,.
\label{eq:kn}
\end{equation}

\medskip

\noindent We note that in Equation~\ref{eq:jstar}, we have adapted our calculation from that of \citet{Romanowsky:2012} by assuming that $v_{2.2_{\rm{C}}} \equiv v_{s}$, where $v_{s}$ is the intrinsic rotation velocity at $2R_{50}$. This assumption is justified since, as discussed in \S~\ref{subsubsec:v_obs}, the two quantities should only differ by a few per cent, on average. The distribution of $j_{*2.2_{\rm{C}}}$ for KGES galaxies with robust kinematics (i.e.\ {\it kinematics} sub-sample galaxies; see \S~\ref{subsec:kinematics_selection}) is shown in Figure~\ref{fig:subsample_survey_comp_histograms}. Their median $j_{*2.2_{\rm{C}}}$ is 441 $\pm$ 43 kpc km s$^{-1}$, with a scatter of $\sigma_{\rm{MAD}} =$ 377~$\pm$ 39~kpc~km~s$^{-1}$. 

\section{Low-redshift Comparison Data and Kinematic Sub-sample Selection} 
\label{sec:lowz_comp_and_subsamples}

To inform and extend our analysis of the KGES galaxies at $z\approx1.5$, we compare their properties to those of star-forming galaxies at $z\approx0.9$ and $z\approx0.04$ with corresponding IFS observations performed respectively as part of KROSS and the SAMI Galaxy Survey. For a fair comparison between redshifts, we match our analysis methods, measurement definitions, and sample selection criteria for galaxies across the three surveys. We also apply beam smearing corrections to the KROSS kinematic measurements in the same manner as outlined for the KGES galaxies in \S~\ref{sec:resolved_properties}. In \S~\ref{subsec:comparison_samples} we provide details of the KROSS and SAMI samples and measurements. In \S~\ref{subsec:kinematics_selection} we describe how we select sub-samples of galaxies with robust kinematics, the ``{\it kinematics}" sub-samples, from each of the three surveys using uniform selection criteria.

\subsection{Comparison samples}
\label{subsec:comparison_samples}

For a comparison sample of star-forming galaxies in the local Universe, we select galaxies from the SAMI Galaxy Survey. We select a ``main sequence" sub-sample for comparison with KGES as the 489 galaxies that are members of the SAMI {\it parent} sub-sample defined in \citet{Tiley:2018}, and with a specific star-formation rate, $\rm{sSFR} \geq 1.5 \times 10^{-11}$ yr$^{-1}$. 

For the SAMI stellar masses, we adopt the values calculated by \citet{Bryant:2015}, derived from $g-i$ colours and $i$ band magnitudes from the Galaxy And Mass Assembly survey \citep[GAMA;][]{Driver:2011} according to the method of \citet{Taylor:2011} and assuming a \citet{Chabrier:2003} IMF. For the SAMI SFRs, we adopt the values measured by \citet{Davies:2016} via the application of the {\sc magphys} SED-fitting routine to extensive, multi-wavelength GAMA photometry for each galaxy.  We adopt the effective (i.e.\ half-light) radii, axial ratios (and corresponding inclinations), and S\'ersic indices measured by \citet{Kelvin:2012} from single component, two-dimensional S\'ersic profile fits to the Sloan Digital Sky Survey \citep{York:2000} $r$ band image for each galaxy. In \citet{Tiley:2018} we calculated the characteristic intrinsic rotation velocity and velocity dispersion for each SAMI galaxy from its spatially-resolved H$\alpha$ and [N {\sc ii}] emission in the same manner as for the KGES galaxies outlined in \S~\ref{sec:resolved_properties}. We adopt those measurements here. Since the ratio of the angular size of the SAMI galaxies to the SAMI PSF is relatively large, the required beam smearing corrections are negligible and therefore omitted. Furthermore, due to the comparatively limited physical size of the SAMI FOV (with respect to the size of the galaxies), the SAMI galaxy velocity dispersions are uniformly calculated from the median of their dispersion maps, rather than their outer dispersion curves. We calculate $j_{*2.2_{\rm{C}}}$ for SAMI galaxies in the same manner as for those in KGES and as described in \S~\ref{subsec:specific_angmom}.

For an additional comparison sample, we select 472 galaxies from the KROSS sample of star-forming galaxies at $z\approx0.9$, spatially-resolved in H$\alpha$ with KMOS and with associated measurements of stellar mass, ionised gas rotation velocity, and velocity dispersion from \citet{Harrison:2017} and half-light radii from \citet{Tiley:2018} (themselves converted to a {\it WMAP9} cosmology from the measurements of \citealt{Harrison:2017}). KROSS galaxies are typical star-forming galaxies for their epochs, the vast majority residing on the main-sequence of star formation for their corresponding redshifts and stellar masses.

For the KROSS galaxies we adopt the stellar masses and star-formation rates calculated and presented in \citet{Harrison:2017}. The former were determined as a function of each galaxy's absolute $H$ band magnitude and the latter from their H$\alpha$ flux, in the same manner as described for the KGES galaxies in \S~\ref{subsec:totfluxes_and_sfrs}, but with a fixed $A_{\rm{H}\alpha,\rm{gas}} = 1.73$ (see  \citealt{Harrison:2017} for further details). We adopt the inclinations and half-light radii for KROSS galaxies presented in the same work, determined respectively via a two-dimensional Gaussian fit and an elliptical aperture curve-of-growth analysis on the highest resolution, and reddest bandpass, image available for each galaxy. S\'ersic indices, measured by \citet{vanderWel:2012} via single component, two-dimensional S\'ersic profile fits to $F$160$W$ ($H$ band) {\it HST} images of galaxies in the CANDELS extragalactic fields, are only available for a sub-set ($\approx$18 per cent) of the resolved KROSS galaxies. We calculate the characteristic intrinsic rotation velocities and velocity dispersions for the KROSS galaxies in the same manner as for KGES galaxies (\S~\ref{sec:resolved_properties}), starting with the $v_{2.2,\rm{obs}}$ and $\sigma_{0,\rm{obs}}$ measured for each KROSS galaxy by \citet{Harrison:2017} from its H$\alpha$ and [N {\sc ii}] emission and applying beam smearing corrections according to Equation~\ref{eq:V22C} and Equation~\ref{eq:sig0C}, respectively. 

We calculate $j_{*2.2_{\rm{C}}}$ for KROSS galaxies in the same manner as for those in KGES and SAMI, with one notable difference. Since we do not have a  measure of $n_{\rm{S}}$ for every KROSS galaxy, we instead assume a fixed $n_{\rm{S}} = 1$ ($k_{n} = 1.19$) for each. This is justified, on average at least, since the median $n_{\rm{S}}$ of those KROSS {\it kinematics} sub-sample galaxies with a measurement (see \S~\ref{subsec:kinematics_selection}) is $1.04 \pm 0.06$, with a scatter of $\sigma_{\rm{MAD}} = 0.42 \pm 0.09$ (consistent with the median and scatter for all KROSS galaxies with a measurement of $n_{\rm{S}}$).  Furthermore, 87 per cent of KROSS {\it kinematics} sub-sample galaxies (86 per cent for KROSS galaxies overall) with a measurement have $n_{\rm{S}} < 2$. For comparison \citep[and as discussed in][]{Harrison:2017}, adopting a fixed $n_{\rm{S}} = 2$ instead would only increase $k_{n}$ (and thus $j_{*2.2_{\rm{C}}}$) by $17$ per cent ($0.07$ dex) compared to $n_{\rm{S}} = 1$, meaning our calculations are anyway robust to our choice of fixed $n_{\rm{S}}$ for KROSS systems. 

\subsection{Kinematics sub-sample selection}
\label{subsec:kinematics_selection}

The final step before proceeding with our analysis is to apply consistent selection criteria to uniformly select galaxies suitable for inclusion in our kinematic analysis in this section i.e. the {\it kinematics} sub-samples. 

For our {\it kinematics} sub-samples, at each redshift we select respectively the 481, 472, and 210 SAMI, KROSS, and KGES galaxies that are main sequence star-forming systems\footnote{We only explicitly select for the main sequence in the SAMI sample (see \S~\ref{subsec:comparison_samples}), the KROSS and KGES systems are only {\it effectively} selected to fall on the main sequence for their epoch.}, spatially-resolved in H$\alpha$ emission, with associated $M_{*}$, $v_{2.2_{\rm{C}}}$, $\sigma_{0_{\rm{C}}}$, and $R_{50}$ measurements (each with corresponding uncertainties), and not flagged as AGN.\footnote{We adopt the \citet{Harrison:2017} AGN flags for the KROSS galaxies. We make no explicit AGN cuts for the SAMI galaxies, except for removing a single system with a very large velocity dispersion ($\sigma_{0_{\rm{C}}} > 500$ km s$^{-1}$), which may be indicative of the presence of an AGN.} We also disregard any galaxies with H$\alpha$ that is insufficiently radially extended to allow for a robust measurement of rotation velocity (see \S~\ref{subsec:halpha_extent}), leaving 420, 457, and 202 SAMI, KROSS and KGES galaxies, respectively. 

To select for galaxies with robust kinematic measurements, we further remove those with a fractional uncertainty in $v_{2.2_{\rm{C}}}$ greater than 30 per cent - leading to remaining sub-samples of 414 SAMI galaxies, 289 KROSS galaxies, and 181 KGES galaxies. Similarly we exclude respectively a further 127, 63, and 55 SAMI, KROSS and KGES galaxies with inclinations outside of the range $45 < i <85$. The lower inclination limit is imposed to remove systems that require a large corresponding correction to their rotation velocity \citep[and are therefore most sensitive to innacuracies in $i$; e.g.][]{Tiley:2016b}. The upper limit excludes galaxies that are very edge on, and thus with an increased probability of suffering from substantial dust obscuration (that in turn may affect the accuracy of properties calculated from their photometry, including $M_{*}$).

As a final step we apply an additional cut in stellar mass to the remaining SAMI galaxies, excluding 98 with $\log_{10}(M_{*}/\rm{M}_{\odot}) < 9$, to match the approximate lower limit of both the KROSS and KGES stellar mass range. 

The 126 KGES galaxies, 226 KROSS galaxies, and 189 SAMI galaxies that remain after application of all the selection criteria listed make up our {\it kinematics} sub-sample at respectively $z\approx1.5$, $\approx0.9$, and $\approx0.04$. 

Histograms of the key galaxy properties for the {\it kinematics} sub-samples are shown in Figure~\ref{fig:subsample_survey_comp_histograms}. They span the same approximate range in $\log_{10}(M_{*}/\rm{M}_{\odot})$ at each redshift (by design) and have median values of $M_{*}$, $R_{50}$, $v_{2.2_{\rm{C}}}$, $v_{2.2_{\rm{C}}}/\sigma_{0_{\rm{C}}}$, and $j_{*2.2_{\rm{C}}}$ that are consistent, after accounting for uncertainties. The only quantities for which the median values differ between the three redshifts are the galaxies' stellar masses, sSFR and $\sigma_{0_{\rm{C}}}$. While the median stellar 
masses and $\sigma_{0_{\rm{C}}}$ increase primarily from $z\approx0.04$ to $z\approx0.9$, and are similar from $z\approx0.9$ to $z\approx1.5$, the 
median sSFR keeps monotonically increasing when moving from SAMI, KROSS to KGES.
It is worth noting that the similarities in sizes between galaxies at the three redshifts investigated here do not necessarily imply a lack of evolution in the mass-size relation of galaxies. On the contrary, they are most likely a result of the selection criteria used to match the sample extracted from SAMI, KROSS and KGES.


\section{The Kinematics of Star-forming Galaxies Over the Past 10 Gyr}
\label{sec:results_and_discussion}

In the previous sections we confirmed that the KGES galaxies at $z\approx1.5$ are ``normal" star-forming systems for their epoch. We also described how we constructed {\it kinematics} sub-samples from KGES, and from comparable IFS surveys of normal star-forming galaxies at lower redshifts, namely KROSS ($z\approx0.9$) and the SAMI Galaxy Survey ($z\approx0.04$), each with matched kinematic measurements and selection criteria. In this section we proceed to compare the kinematics of galaxies in these sub-samples at $z\approx1.5$, $\approx0.9$, and $\approx0.04$. Our aim is to determine how prevalent disc-like characteristics are within the star-forming population over the past $\approx$10 Gyr, and to measure to what extent the angular momentum content of star-forming galaxies has varied over the same period. 

\subsection{Disc-like characteristics of star-forming galaxies}
\label{subsec:disciness}

As explained in \S~\ref{sec:intro}, whether or not a galaxy exhibits a disc structure (either in stars or gas) should be intimately linked to its history of assembly, including its initial formation and subsequent evolution. 

A galaxy may be deemed to host a pure disc if its S\'ersic index ($n_{\rm{S}}$, measured from its stellar light) is consistent with unity, i.e.\ that of an exponential disc. \citet{Gillman:2020} measured $n_{\rm{S}}$ for each KGES galaxy, and a similar measurement is available for each SAMI galaxy via modelling of its $r$ band image \citep{Kelvin:2012}. However, since a measure of $n_{\rm{S}}$ is only available for a sub-set of KROSS galaxies (see \citealt{Harrison:2017} for further details), we avoid an extended comparison of $n_{\rm{S}}$ for galaxies in our {\it kinematics} sub-samples at each redshift. We simply note that the median $n_{\rm{S}}$ for those galaxies with a measurement at $z\approx1.5$ and $z\approx0.9$ is consistent with unity ($1.0 \pm 0.2$ and $1.04 \pm 0.06$, respectively). The median $n_{\rm{S}}$ for those galaxies at $z\approx0.04$ in our analysis with a measurement is $1.17 \pm 0.05$. 

For alternative indicators of how disc-like our galaxies are, we also examine their kinematic properties. We follow the example of \citet{Tiley:2018} who used the ratio of galaxies' rotation-to-dispersion ($v/\sigma$; a global proxy for how rotation-dominated a galaxy's kinematics are, or similarly how closely the galaxy obeys circular motion), and the extent to which their velocity field resembles that of a disc to quantify their ``disciness". The latter is determined via the $R^{2}_{\rm{disc}}$ goodness-of-fit parameter, calculated from the residuals between a galaxy's observed velocity field and the corresponding best fitting disc model velocity field. The $R^{2}_{\rm{disc}}$ value describes the extent to which the total variation in a galaxy's $v_{\rm{obs}}$ map is explained by the best fitting model map, varying from 0 (not described by the model at all) to 1 (completely described by the model). 

For the former we adopt the quantity $v_{2.2_{\rm{C}}}/\sigma_{0_{\rm{C}}}$. The ($\log_{10}$) distributions of $v_{2.2_{\rm{C}}}/\sigma_{0_{\rm{C}}}$ for our {\it kinematics} sub-sample galaxies at each redshift are shown in Figure~\ref{fig:subsample_survey_comp_histograms}. The median $v_{2.2_{\rm{C}}}/\sigma_{0_{\rm{C}}}$ is $2.5 \pm 0.2$, $2.6 \pm 0.2$, and $3.0 \pm 0.1$ for galaxies at $z\approx1.5$, $\approx0.9$, and $\approx0.04$, respectively. The corresponding scatters are $\sigma_{\rm{MAD}} = 1.6 \pm 0.2$, $1.8 \pm 0.2$, and $1.2 \pm 0.1$. Despite the large scatters at high redshift, our {\it kinematics} sub-sample galaxies have similar average ratios of rotation-to-dispersion support in their (gas) kinematics, being {\it formally} rotation-dominated ($v_{2.2_{\rm{C}}}/\sigma_{0_{\rm{C}}} > 1$) on average at every redshift. \citet{Tiley:2018} discuss how a limit of $v/\sigma = 3$ is more appropriate for determining whether a galaxy's kinematics are {\it truly} rotation-dominated since, under sensible assumptions \citep[e.g.][]{Kormendy:2001}, ratios above this limit ensure that the rotation velocity term in the collisionless Boltzmann equation accounts for at least 90 per cent of the galaxy's dynamical mass. Accounting for uncertainties, the median $v_{2.2_{\rm{C}}}/\sigma_{0_{\rm{C}}}$ for {\it kinematics} sub-sample galaxies does not significantly differ from this alternative limit at any of the three redshift considered in our analysis.

For each galaxy in the KROSS and SAMI {\it kinematics} sub-samples we adopt measurements of $R^{2}_{\rm{disc}}$ from \citet{Tiley:2018}, determined by fitting a two-dimensional extension of the disk model described in Equation~\ref{eq:diskmodel} to each galaxy' $v_{\rm{obs}}$ map and examining the resultant residuals. We apply the same analysis to the KGES galaxies, calculating $R^{2}_{\rm{disc}}$ for each H$\alpha$-resolved system. The median $R^{2}_{\rm{disc}}$ for KGES, KROSS, and SAMI {\it kinematics} sub-sample galaxies is respectively $0.78 \pm 0.03$, $0.86 \pm 0.01$, and $0.919 \pm 0.009$, with corresponding scatters of $\sigma_{\rm{MAD}} = 0.19 \pm 0.04$, $0.14 \pm 0.01$, and $0.07 \pm 0.01$. Whilst the median $R^{2}_{\rm{disc}}$ does decrease with increasing redshift, within uncertainties it is still consistent at each epoch with being equal to or above the lower limit of $R^{2}_{\rm{disc}} = 0.8$ used by \citet{Tiley:2018} previously to define, in part, ``discy" galaxies at $z\approx0.9$ and $\approx0.04$.

\subsubsection{Disc fraction as a function of redshift}
\label{subsubsec:disc_fractions}

\begin{figure*}
\centering
\includegraphics[width=.49\textwidth]{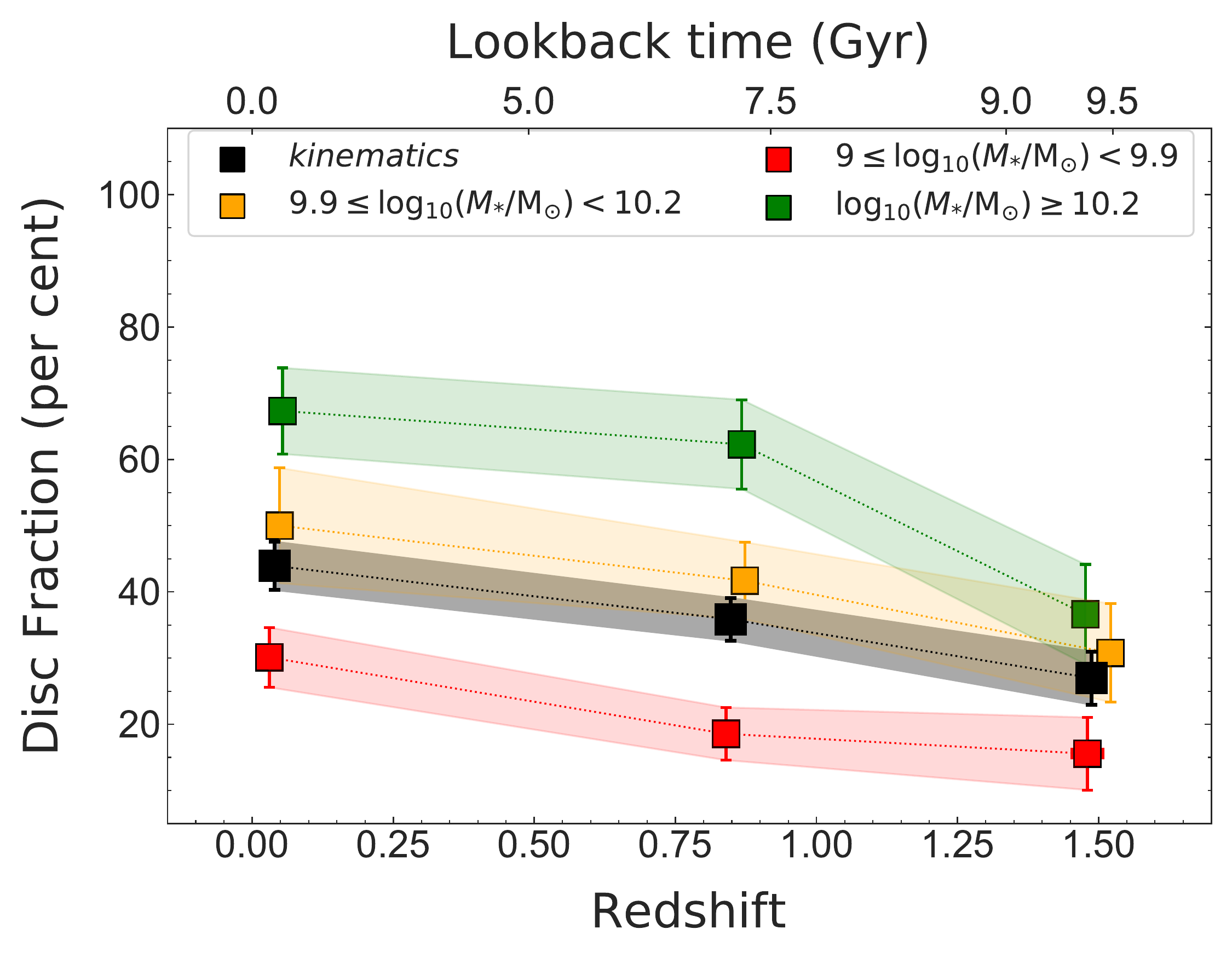}
\includegraphics[width=.49\textwidth]{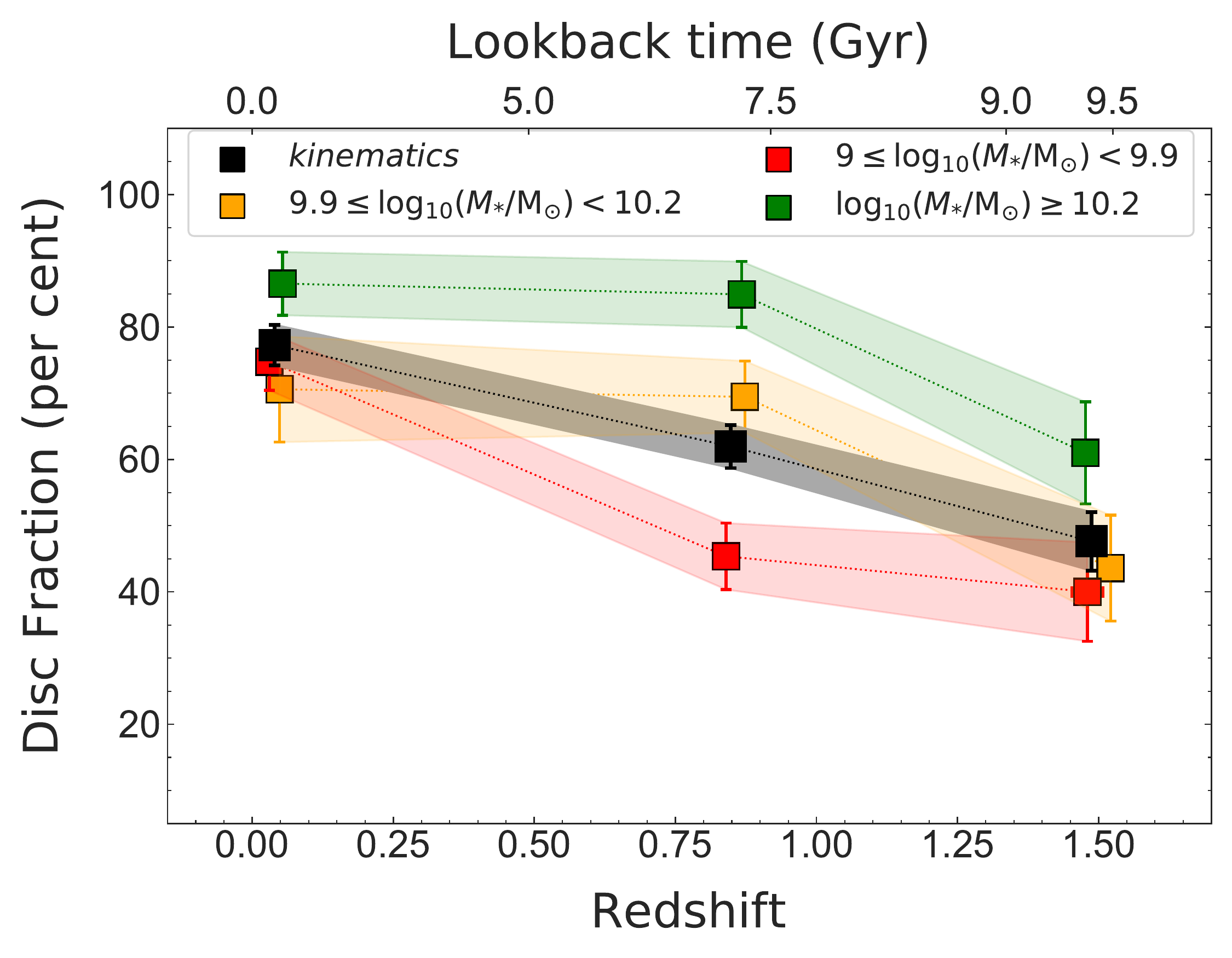}
\caption{%
The fraction of discs amongst our {\it kinematics} sub-sample galaxies as a function of redshift (black squares), and also within bins of stellar mass (coloured squares). Discs as defined as galaxies with $R^{2}_{\rm{disc}} > 0.8$ and either $v_{2.2_{\rm{C}}}/\sigma_{0_{\rm{C}}} > 3$ (left) or $v_{2.2_{\rm{C}}}/\sigma_{0_{\rm{C}}} > 1$ (right). To guide the eye, we interpolate between our measurements at each redshift (dashed lines in corresponding colours) and their corresponding uncertainty (filled regions in corresponding colours). Overall, the disc fraction is a much stronger function of galaxies' stellar masses than their redshifts. 
\label{fig:disc_fractions}}
\end{figure*}

Considering the {\it individual} metrics of ``disciness" adopted in this work, on average massive, star-forming galaxies exhibit disc-like characteristics to a comparable degree regardless of whether they reside at $z\approx$1.5, $\approx$0.9, or $\approx$0.04. Despite this, we still need to quantify exactly how the prevalence of disc systems amongst the star-forming population has changed since $z\approx1.5$. Based closely on the criteria adopted by \citet{Tiley:2018}, we identify disc galaxies in our analysis as those that satisfy the {\it combined} criteria that their $v_{2.2_{\rm{C}}}/\sigma_{0_{\rm{C}}} > 3$ and $R^{2}_{\rm{disc}} > 0.8$. Interestingly, for all three samples the $v_{2.2_{\rm{C}}}/\sigma_{0_{\rm{C}}}$ ratio is the more restrictive criteria in the selection of disc galaxies, but the difference decreases significantly with increasing redshift. For example, while for SAMI 77\% of the sample fulfills the $R^{2}_{\rm{disc}}$ criteria and only 49\% the one based $v_{2.2_{\rm{C}}}/\sigma_{0_{\rm{C}}}$, for KGES these fractions drop to 37\% and 48\%, respectively.

In Figure~\ref{fig:disc_fractions} (left), we plot the fraction of disc galaxies within the {\it kinematics} sub-samples, as a function of redshift and stellar mass. For {\it kinematics} sub-sample galaxies at $z\approx1.5$, $z\approx0.9$, and $z\approx0.04$ we calculate disc fractions (and corresponding bootstrapped $1\sigma$ uncertainties) of respectively $27 \pm 4$, $36 \pm 3$, and $44 \pm 4$ per cent. There is a small systematic increase in the disc fraction with decreasing redshift. This increase is not statistically significant between individual adjacent redshift bins. However, there is a significant difference in the disc fraction between the highest and lowest redshift bins, i.e.\ between {\it kinematics} sub-sample galaxies at $z\approx1.5$ and $z\approx0.04$. Nevertheless, although statistically significant ($3.1\sigma$), this difference is only modest. 

Next we consider the disc fraction at each redshift within three bins of increasing stellar mass. Within the lowest and intermediate mass bins the disc fraction does not significantly differ between {\it kinematics} sub-sample galaxies in any of the three redshift bins. For the highest mass bin, the disc fractions at $z\approx0.9$ and $z\approx0.04$ are consistent with one another. However, in the same bin, there is respectively a marginally significant ($2.6\sigma$) and significant ($3.0\sigma$) difference between the disc fraction of {\it kinematics} sub-sample galaxies at $z\approx1.5$ and of those at $z\approx0.9$ and $z\approx0.04$ (galaxies at $z\approx1.5$ have a lower disc fraction in each case). Importantly, the difference in the disc fraction between {\it stellar mass bins} at fixed {\it redshift} is comparable to, or larger than, the difference we measure between redshifts at fixed stellar mass. For example, the disc fraction for galaxies in the highest mass bin is significantly larger than those in the lowest mass bin at both $z\approx0.9$ and $z\approx0.04$ (respectively a $5.6\sigma$ and $4.6\sigma$ difference). We see a similar, marginally significant ($2.3\sigma$) difference between the same stellar mass bins for galaxies at $z\approx1.5$. As extensively discussed in \S~7.3, our lower stellar mass bins in KROSS and KGES are biased towards galaxies above the main sequence. Thus, we cannot exclude that this selection bias might be behind the marginal difference in redshift evolution of the disc fraction for low and high stellar mass galaxies.

We also note that our calculated disc fractions are in general lower than those determined for star-forming galaxies at $z\approx1$ and $z\approx2$ by \citet{Wisnioski:2015} - respectively $70$--$90$ and $47$--$74$ per cent. These are based on a number of criteria including the appearance of the galaxy's velocity map, whether the galaxy satisfies $v/\sigma > 1$, the extent of any misalignment between its kinematic and photometric position angles, and whether its kinematic centre is spatially coincident with respectively the peak of its continuum emission and its peak velocity dispersion. The range at each redshift corresponds to how many or few of the criteria are implemented. Our estimates are more consistent with those of \citet{Rodrigues:2017}, who use the same criteria as \citet{Wisnioski:2015} but come to the alternative conclusion that only 53 per cent of massive, star-forming galaxies at $z\approx1$ exhibit disc structures. Although, their estimate reduces to 25 per cent when they introduce a further criterion based on the visual morphology of galaxies. 

It's clear that the {\it absolute} value of the disc fraction is highly sensitive to the choice of criteria used to identify discs, as well as the implementation of those criteria if there is any subjectivity associated with them. Thus we refrain from any attempts to justify in detail the difference between our estimates of the disc fractions of star-forming galaxies in our sample and those measured previously for galaxies at similar stellar masses and redshifts, given we adopt metrics of disciness that are different again to each of the two studies discussed. To make this point even clearer, in the right panel of Figure~\ref{fig:disc_fractions} we show how our disc fractions do change if we use the less stringent cut of $v/\sigma > 1$, while keeping $R^{2}_{\rm{disc}} > 0.8$. It is clear that the overall fractions are now consistent with the values presented by \citet{Wisnioski:2015}, but the mass and redshift trends remain the same.

Indeed, we stress here the most important point that, when we apply uniform criteria to identifying discs, there is at most only modest (significant) differences in the disc fraction {\it between redshifts}. And that this remains true if we also consider galaxies within the same stellar mass bin at each redshift. The disc fraction does, however, significantly differ {\it between mass bins} in many cases across the three redshifts, with the magnitude of the difference as large or larger than that we measure between redshifts. 

Many previous studies have shown that star-formation has predominantly taken place within disc structures in galaxies throughout cosmic history, as the atomic and molecular gas that feeds star-formation is dissipative and thus prone to settling into a disc. The fact that we find a similar disc fraction at each redshift then is perhaps unsurprising given that we have explicitly selected for star-forming galaxies in each case, which by association are those most likely to host discs. Nevertheless, it need not be the case that the disc fraction in the high stellar mass star-forming population only modestly increases over a period of 10 Gyr. The fact we generally see only small differences at fixed stellar mass between $z\approx1.5$ and $z\approx0.04$ might suggest that star-forming galaxies of a given stellar mass tend to form via similar formation pathways, regardless of the cosmic epoch. 

The systematic increase in disc fraction with increasing stellar mass at fixed redshift (across all three redshift bins) is also in line with previous studies \citep[e.g.][]{Kassin:2012,vanderWel:2014,Simons:2016,Simons:2017,Johnson:2018,Wisnioski:2019} that report evidence for ``kinematic downsizing" amongst the star-forming population at $z \lesssim 2$. In this scenario star-forming galaxies generally grow in a hierarchical fashion, evolving from disordered to ordered systems as their gas settles down to form discs. The most massive galaxies at any epoch have formed a larger fraction of their stars at earlier times \citep[i.e.\ conventional ``downsizing";][]{Cowie:1996} and, given their larger mass, are more stable to disruptive processes such as minor mergers, or gas inflows or outflows (and/or may undergo them less often). Hence, when we focus on gas kinematic, the most massive star-forming galaxies tend to be more kinematically ``mature", i.e.\ more disc-like, than lower mass systems at any given epoch. However, we cannot rule out the possibility that the systematic increase in disc fraction with increasing stellar mass that we observe may instead be the result, or the partial result, of an aperture effect. For example, it is possible that our rotation velocity measurement, $v_{2.2_{\rm{C}}}$, is a systematically increasing fraction of the ``maximum" rotation velocity with increasing stellar mass of a galaxy. Similarly, nor should we ignore the fact that $v_{2.2_{\rm{C}}}$ and $\sigma_{0_{\rm{C}}}$, respectively the numerator and denominator in the ratio $v_{2.2_{\rm{C}}}/\sigma_{0_{\rm{C}}}$ that we use as one of our two indicators of whether a galaxy is a disc, have different dependencies on stellar mass. Indeed the former correlates more strongly with stellar mass than the latter for our {\it kinematics} sub-sample galaxies at each redshift. Thus, it is also possible that the correlation between disc fraction and stellar mass at fixed redshift is also driven, to some extent, by the differing stellar mass dependencies of $v_{2.2_{\rm{C}}}$ and $\sigma_{0_{\rm{C}}}$.

\subsection{Specific angular momentum of star-forming galaxies}
\label{subsec:angmom}

\begin{figure*}
\begin{minipage}[]{1.0\textwidth}\
\includegraphics[width=1.\textwidth,trim= 10 0 90 0,clip=True]{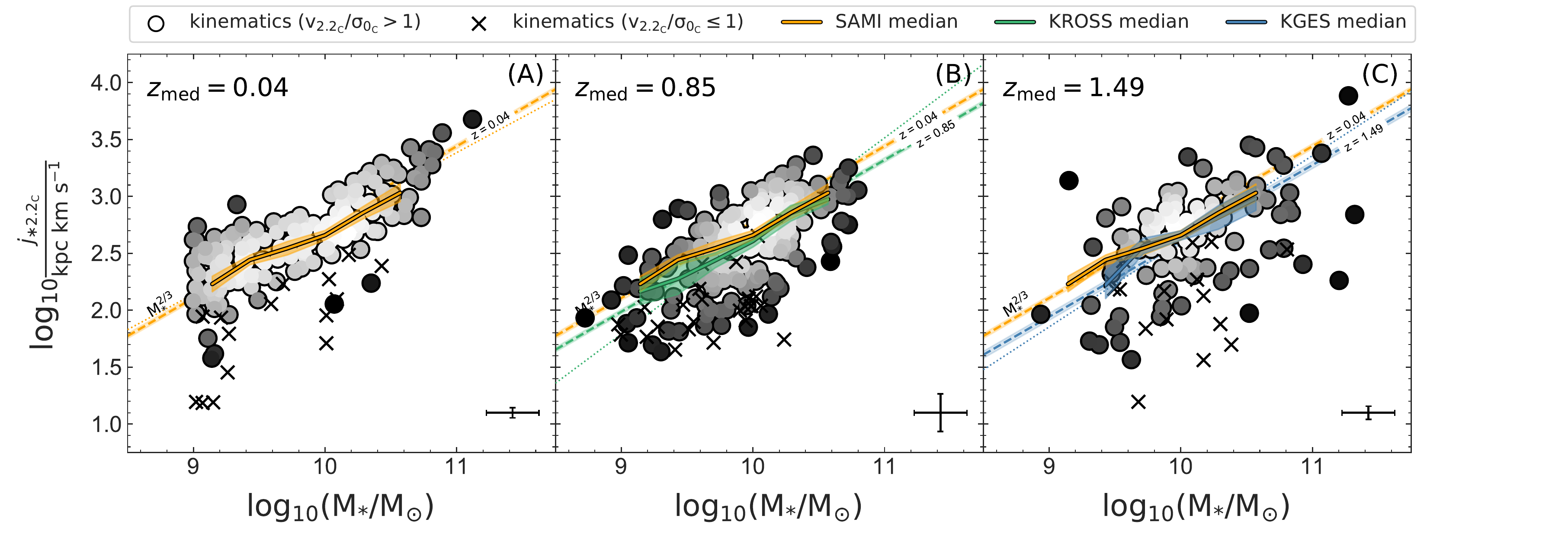}
\end{minipage}
\caption{%
The specific angular momenta of star-forming galaxies in the SAMI (A, left), KROSS (B, middle) and KGES (C, right) {\it kinematics} sub-samples, as a function of their stellar masses. The galaxies span three different redshift slices at $z \approx 0.04$, $\approx0.9$, and $\approx1.5$. Scatter points are colour-coded by the local spatial density of the points themselves (dark to light, low to high density; quantified via a Gaussian kernel density estimate and {\sc loess}-smoothed). The best linear fit (with slope fixed to $2/3$) and its $1\sigma$ uncertainty envelope is shown for the SAMI galaxies as an orange dashed line and filled region respectively in each panel. The corresponding best fits to the KROSS and KGES galaxies are shown as respectively a green and a blue dashed line and filled region in the middle and right panels. Similarly the median trends, and their $1\sigma$ uncertainty envelopes, for each sample are shown by solid lines and filled regions in the same colours. There is a small offset between the normalisation of the $z=0.04$ best fit linear trend, and that for the higher redshifts galaxies. However, the median trends do not significantly differ as a function of redshift. Thus, on median average, at fixed stellar mass a massive, star-forming galaxy has the same specific angular momentum regardless of whether it resides at $z\approx0$, $\approx0.9$, or $\approx1.5$. %
     }%
\label{fig:jstar_vs_Mstar}
\end{figure*}


\subsubsection{Best-fitting linear trends}
\label{subsubsec:linear trends}

In Figure~\ref{fig:jstar_vs_Mstar}, we plot the $j_{*2.2_{\rm{C}}}$ of the {\it kinematics} sub-sample galaxies as a function of their stellar masses at $z\approx1.5$, $\approx0.9$, and  $\approx0.04$, showing that the two quantities are correlated with one another at each of the three redshifts. 

Most galaxy scaling relations, including the $j_{*2.2_{\rm{C}}}$--$M_{*}$ relation, are usually assumed to follow a linear relation in log-space (i.e.\ a power law), and thus are traditionally modelled accordingly with a straight line. However, visual inspection of the positions of our {\it kinematics} sub-sample galaxies in the $j_{*2.2_{\rm{C}}}$--$M_{*}$ plane reveals that this approach may not always be the best approach. Firstly, outlying galaxies are preferentially scattered toward low $j_{*2.2_{\rm{C}}}$ at fixed $M_{*}$ (i.e. the scatter is not symmetrical about the main locus of scatter points) at all three redshifts, and that this is true even after excluding dispersion-dominated systems ($v_{2.2_{\rm{C}}}/\sigma_{0_{\rm{C}}} \leq 1$) at each epoch. Such a skewed scatter can bias the best fitting linear normalisation, and may also affect the best fit slope if the magnitude of the scatter also depends on $M_{*}$ -- as is the case at least for the KROSS galaxies at $z\approx0.9$. Furthermore, irrespective of the scatter, it is not clear, visually, that the slope of the $j_{*2.2_{\rm{C}}}$--$M_{*}$ relation is constant with stellar mass at each redshift, particularly for galaxies in our analysis at $z\approx0.04$. 
Thus, it is useful to combine a simple linear fit with an estimate of the median trends without any assumptions onf the functional shape of the relation.

For ease of comparison with past and future studies, we start by fitting two versions of a linear relation to the positions of our {\it kinematics} sub-sample galaxies in the $j_{*2.2_{\rm{C}}}$--$M_{*}$ plane at each redshift - the first with a slope that is free to vary, and the second with a slope fixed to a value of our choosing. 

We use 
the {\sc hyperfit} package \citep{Robotham:2015} to find the best linear fit (minimising the orthogonal scatter) to the $j_{*2.2_{\rm{C}}}$--$M_{*}$ relation for rotation-dominated ($v_{2.2_{\rm{C}}}/\sigma_{0_{\rm{C}}} > 1$) galaxies at each redshift. In performing the fit we ignore dispersion-dominated galaxies that, although in the minority, may bias the best-fit parameters since they tend to be outlying from the main locus of points in the $j_{*2.2_{\rm{C}}}$--$M_{*}$ plane at each redshift (although see later for comments on the median trends in this regard). The linear fit takes the form $\log_{10}(j_{*2.2_{\rm{C}}}/\rm{kpc}\ \rm{km}\ \rm{s}^{-1}) =  \alpha [ \log_{10}( M_{*}/\rm{M}_{\odot})-10] + \beta$. 

The best fitting parameters for the linear fits are listed in Table~\ref{tab:bestfit_lines}, and the resultant trends are included in Figure~\ref{fig:jstar_vs_Mstar}. Within uncertainties, the slopes of the relation at $z\approx0.04$ and $z\approx1.5$ are consistent with one another, and with the expectation from tidal torque theory of $\alpha = 2/3$ \citep[e.g.][]{Catelan:1996}. Such a slope should arise in the case that baryons and dark matter are well mixed in ``proto-galaxies" (i.e.\ in haloes before the initial baryonic collapse), and that after decoupling the baryons subsequently retain their initial angular momentum so that they mirror the relationship expected between the angular momentum and mass of the dark matter, i.e.\ $j \propto M^{2/3}$. Any significant deviation from $\alpha = 2/3$, as in the case of the best fitting slope for {\it kinematics} sub-sample galaxies at $z\approx0.9$, could imply a mass dependence on the transfer of halo angular momentum to baryonic angular momentum, or subsequent retention of the latter. However, in this particular case, the steeper slope at $z\approx0.9$ is likely biased due to subtle selection effects at low stellar masses in the H$\alpha$-selected samples in the higher redshift bins. This is discussed in more detail in \S~\ref{subsec:galaxy_halo_link}, where we more explicitly explore the link between baryonic and halo angular momenta of galaxies in our sample.

\begin{table}
\caption{Parameters of the best-fitting straight lines to the $j_{*2.2_{\rm{C}}}$--$M_{*}$ relations for rotation-dominated ($v_{2.2_{\rm{C}}}/\sigma_{0_{\rm{C}}} > 1$) {\it kinematics} sub-sample galaxies at each redshift shown in Figure~\ref{fig:jstar_vs_Mstar}.} \label{tab:bestfit_lines}
\centering
\begin{tabular}{ lccc}
\hline
Fit & Median Redshift & $\alpha$ & $\beta$  \\
\hline
Free & 0.04 & 0.62 $\pm$ 0.03 & 2.76 $\pm$ 0.02  \\
slope & 0.85 & 0.86 $\pm$ 0.06 & 2.65 $\pm$ 0.02  \\
 & 1.49 & 0.75 $\pm$ 0.11 & 2.61 $\pm$ 0.04  \\\hline
Fixed & 0.04 & \phantom{00000.} 2/3 \phantom{$\pm$ 0.02} & 2.77 $\pm$ 0.02  \\
slope & 0.85 & \phantom{00000.} 2/3 \phantom{$\pm$ 0.02} & 2.65 $\pm$ 0.02  \\
 & 1.49 & \phantom{00000.} 2/3 \phantom{$\pm$ 0.02} & 2.61 $\pm$ 0.03  \\\hline
\end{tabular}
\end{table}

\begin{table}
\caption{Median specific angular momentum ($j_{*2.2_{\rm{C}}}$) per bin of stellar mass for the rotation-dominated ($v_{2.2_{\rm{C}}}/\sigma_{0_{\rm{C}}} > 1$) {\it kinematics} sub-sample galaxies at each redshift shown in Figure~\ref{fig:jstar_vs_Mstar}.} \label{tab:median}
\centering
\begin{tabular}{ lccc}
\hline
 $\log M_{*}/M_{\odot}$ &  \multicolumn{3}{c}{$\log j_{*2.2_{\rm{C}}}/(\rm kpc~km~s^{-1})$}  \\
              &    SAMI &  KROSS & KGES \\
              &    $z_{med}$=0.04    & $z_{med}$=0.85 & $z_{med}$=1.49 \\
\hline              
9.14          &      2.23$\pm$0.07   &    2.16$\pm$0.12           &       --          \\
9.43          &      2.44$\pm$0.04   &    2.28$\pm$0.14           &      2.22$\pm$0.12   \\
9.71          &      2.55$\pm$0.06   &    2.44$\pm$0.07           &      2.56$\pm$0.07       \\
10.00          &     2.66$\pm$0.04   &    2.61$\pm$0.05           &      2.66$\pm$0.04     \\
10.29          &     2.86$\pm$0.05   &    2.85$\pm$0.07           &      2.82$\pm$0.10       \\
10.57          &     3.03$\pm$0.08   &    2.97$\pm$0.04           &      2.98$\pm$0.11      \\
\hline

\end{tabular}
\end{table}

\begin{figure*}
\centering
\begin{minipage}[]{1.01\textwidth}
\centering
\includegraphics[width=.925\textwidth,trim= 10 0 80 0,clip=True]{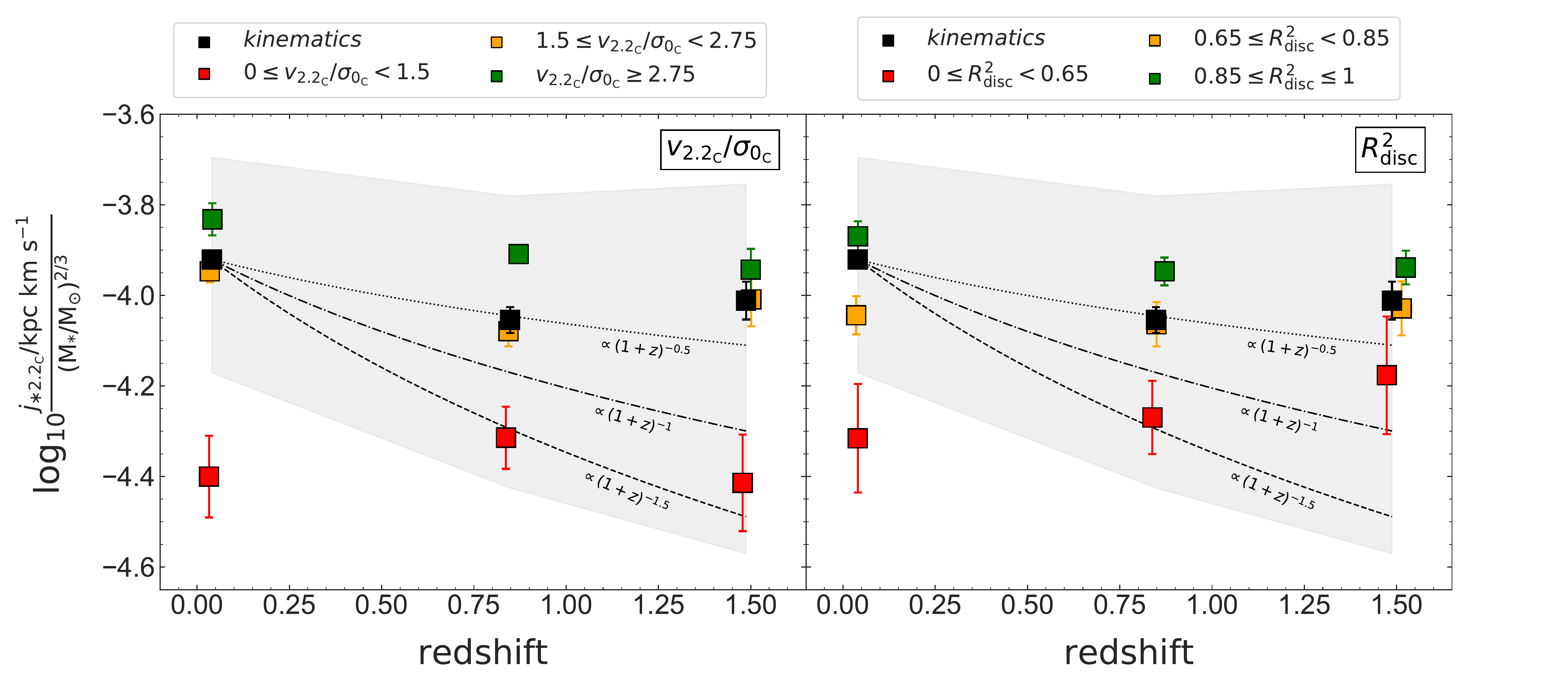}
\end{minipage}
\caption{%
The median normalisation (assuming $j_{*2.2_{\rm{C}}} \propto M_{*}^{2/3}$) of {\it kinematics} sub-sample galaxies in the specific angular momentum-stellar mass plane, as a function of their redshifts and ``disciness". The latter is quantified in the left panel by the galaxies' (gas) rotation velocity-to-velocity dispersion ratios ($v_{2.2_{\rm{C}}}/\sigma_{0_{\rm{C}}}$), and in the right panel by how well their H$\alpha$ velocity field is described by a two-dimensional exponential disc model ($R^{2}_{\rm{disc}}$; the fraction of the variation in the velocity field that is explained by the best fit two dimensional disc model). The total median normalisation at each redshift is shown by the black squares. The median normalisations at each redshift in bins of disciness are shown by the coloured squares. The bin boundaries are given in the legend of each panel. The $1\sigma$ scatter envelope (spanning from the 16$^{\rm{th}}$ to the 84$^{\rm{th}}$ percentile) for the normalisations at each redshift is shown as an underlying filled grey region. We linearly interpolate this between redshifts to better highlight any changes in the scatter. We also overlay tracks of $j_{*2.2_{\rm{C}}}/M_{*}^{2/3} \propto (1 + z)^{-n}$, with $n = 0.5$, $1$, and $1.5$. The normalisation of the specific angular momentum-stellar mass relation does not evolve significantly with redshift. It is instead a strong function of a galaxy's disciness, which is uniform across all three redshift bins.   %
     }%
\label{fig:jstarMstar23_redshift}
\end{figure*}

To measure whether the normalisation of the $j_{*2.2_{\rm{C}}}$--$M_{*}$ relation changes between redshifts, we also find the best {\it fixed-slope} linear fit to the relation for rotation-dominated galaxies at each epoch. For simplicity, we fix $\alpha = 2/3$ since two out of three of the free fit slopes are consistent with this theoretical expectation. The best fitting parameters from the fixed-slope fits are listed in Table~\ref{tab:bestfit_lines}.  We find small, but statistically significant, offsets between the normalisation of the best fixed slope linear fit between redshifts; at fixed $M_{*}$, the average $j_{2.2_{\rm{C}}}$ of the rotation-dominated star-forming galaxies in our {\it kinematics} sub-samples increases by $0.12 \pm 0.03$ dex ($32^{+9}_{-9}$ per cent) between $z\approx0.9$ and $z\approx0.04$, and $0.16 \pm 0.04$ dex ($45^{+14}_{-13}$ per cent) between $z\approx1.5$ and $z\approx0.04$. These differences in normalisation are slightly smaller than, but still qualitatively consistent with, the results of previous studies of the $j_{*2.2_{\rm{C}}}$--$M_{*}$ for star-forming galaxies at similar redshifts. For example \citet{Harrison:2017} find that $z\approx0.9$ star-forming galaxies are offset lower by $\approx0.2$--$0.3$ dex in the $j_{*}$--$M_{*}$ plane at fixed stellar mass compared to $z = 0$ spiral galaxies, and \citet{Swinbank:2017} find a similar offset of $\approx0.2$ dex for star-forming galaxies at $z\approx1$ and Sb or Sc galaxies in the local Universe.

\subsubsection{Median trends}
\label{subsubsec:median_trends}

The median $j_{*2.2_{\rm{C}}}$--$M_{*}$ trends for all {\it kinematics} sub-sample galaxies at each redshift (i.e.\ both rotation- and dispersion-dominated systems) are shown in Figure~\ref{fig:jstar_vs_Mstar}, and presented in Table~\ref{tab:median}. Firstly, we note that, after accounting for uncertainties, each of the median trends is consistent with a straight line with $\alpha = 2/3$ that intersects the trend line at $\log_{10}(M_{*}/\rm{M}_{\odot}) = 10$ (with the exception of one mass bin in the $z\approx0.04$ median trend). Secondly, we find no significant difference in median $j_{*2.2_{\rm{C}}}$ between redshifts in any mass bin. In other words, and despite indications to the contrary from the linear fits discussed in the previous section, on (median) average, massive star-forming galaxies broadly obey a $j_{*2.2_{\rm{C}}} \propto M_{*}^{2/3}$ proportionality at $z\approx1.5$, $z\approx0.9$, or $z\approx0.04$. And, at fixed stellar mass, a typical massive star-forming galaxy at $z \lesssim 1.5$ has the same $j_{*2.2_{\rm{C}}}$, regardless of its redshift. 

To investigate the latter point further, we also explicitly examine to what extent the normalisation, which we define as $j_{*2.2_{\rm{C}}}/M_{*}^{2/3}$ (assuming $j_{*2.2_{\rm{C}}} \propto M_{*}^{2/3}$ after examination of the median trends), depends on other properties that we have measured.  Figure~\ref{fig:jstarMstar23_redshift} shows the median normalisation of our {\it kinematics} sub-sample galaxies in bins of redshift, $v_{2.2_{\rm{C}}}/\sigma_{0_{\rm{C}}}$, and $R^{2}_{\rm{disc}}$. We note that we also examined the dependence of the normalisation on $M_{*}$, $R_{50}$, $v_{2.2_{\rm{C}}}$, and $\sigma_{0_{\rm{C}}}$. However, ignoring $R_{50}$ and $v_{2.2_{\rm{C}}}$ which both linearly correlate with $j_{*2.2_{\rm{C}}}$ by definition of the latter, we find the normalisation to depend mostly strong on $v_{2.2_{\rm{C}}}/\sigma_{0_{\rm{C}}}$ and $R^{2}_{\rm{disc}}$. We therefore focus solely on these two quantities in our analysis. 

In line with the general trend seen in the running medians in Figure~\ref{fig:jstar_vs_Mstar}, from Figure~\ref{fig:jstarMstar23_redshift} it is clear that the median normalisation of the $j_{*2.2_{\rm{C}}}$--$M_{*}$ relation for individual massive star-forming galaxies that comprise our {\it kinematics} sub-samples changes very little as a function of redshift (increasing by $0.13 \pm 0.03$ and $0.09 \pm 0.05$ dex from respectively $z\approx0.9$ and $z\approx1.5$ to $z\approx0.04$, and decreasing by $0.04 \pm 0.05$ dex from $z\approx1.5$ to $z\approx0.9$). In most cases these differences are not statistically significant, with the exception of the difference between $z\approx0.9$ and $z\approx0.04$ ($4.3\sigma$), which is nevertheless only modest in magnitude. The normalisation is instead a much strong function of either $v_{2.2_{\rm{C}}}/\sigma_{0_{\rm{C}}}$ or $R^{2}_{\rm{disc}}$ i.e.\ how disc-like a galaxy's (gas) kinematics are, and this dependence appears approximately uniform across each of the three redshifts we consider. In other words, within bins of either $v_{2.2_{\rm{C}}}/\sigma_{0_{\rm{C}}}$ or $R^{2}_{\rm{disc}}$ the normalisation is constant with redshift, but it deviates strongly {\it between} bins (differing by $\approx0.4$--$0.6$ dex and $0.2$--$0.4$ dex between the lowest and highest bins of respectively $v_{2.2_{\rm{C}}}/\sigma_{0_{\rm{C}}}$ or $R^{2}_{\rm{disc}}$ at fixed redshift, with the exact difference depending on the redshift bin itself). 

Upon first consideration, we should be cautious of physically interpreting a correlation between $j_{*2.2_{\rm{C}}}/M_{*}^{2/3}$ and $v_{2.2_{\rm{C}}}/\sigma_{0_{\rm{C}}}$. For instance, if $\sigma_{0_{\rm{C}}}$ is relatively constant across the sample at each epoch, then the link between $v_{2.2_{\rm{C}}}/\sigma_{0_{\rm{C}}}$ and the $j_{*2.2_{\rm{C}}}$--$M_{*}$ normalisation may simply reflect the fact that $j_{*2.2_{\rm{C}}}$ linearly correlates with $v_{2.2_{\rm{C}}}$. Of course, in reality, the galaxies in our analysis at each redshift exhibit a range of $\sigma_{0_{\rm{C}}}$, the individual values of which may (or may not) also depend on other galaxy properties, including $v_{2.2_{\rm{C}}}$ itself. So the picture is likely not that simple. Nevertheless, it is reassuring that we also see a similar trend if we instead consider $R^{2}_{\rm{disc}}$, which is an independent measure of disciness that does not incorporate any quantities used to calculate $j_{*2.2_{\rm{C}}}$. We therefore conclude that the $j_{*2.2_{\rm{C}}}$--$M_{*}$ normalisation does not differ between redshifts for galaxies that are equally disc-like -- at least in terms of their gas kinematics. 

\begin{figure*}
\centering
\begin{minipage}[]{1.01\textwidth}
\centering
\includegraphics[width=1.01\textwidth,trim= -10 0 0 0,clip=True]{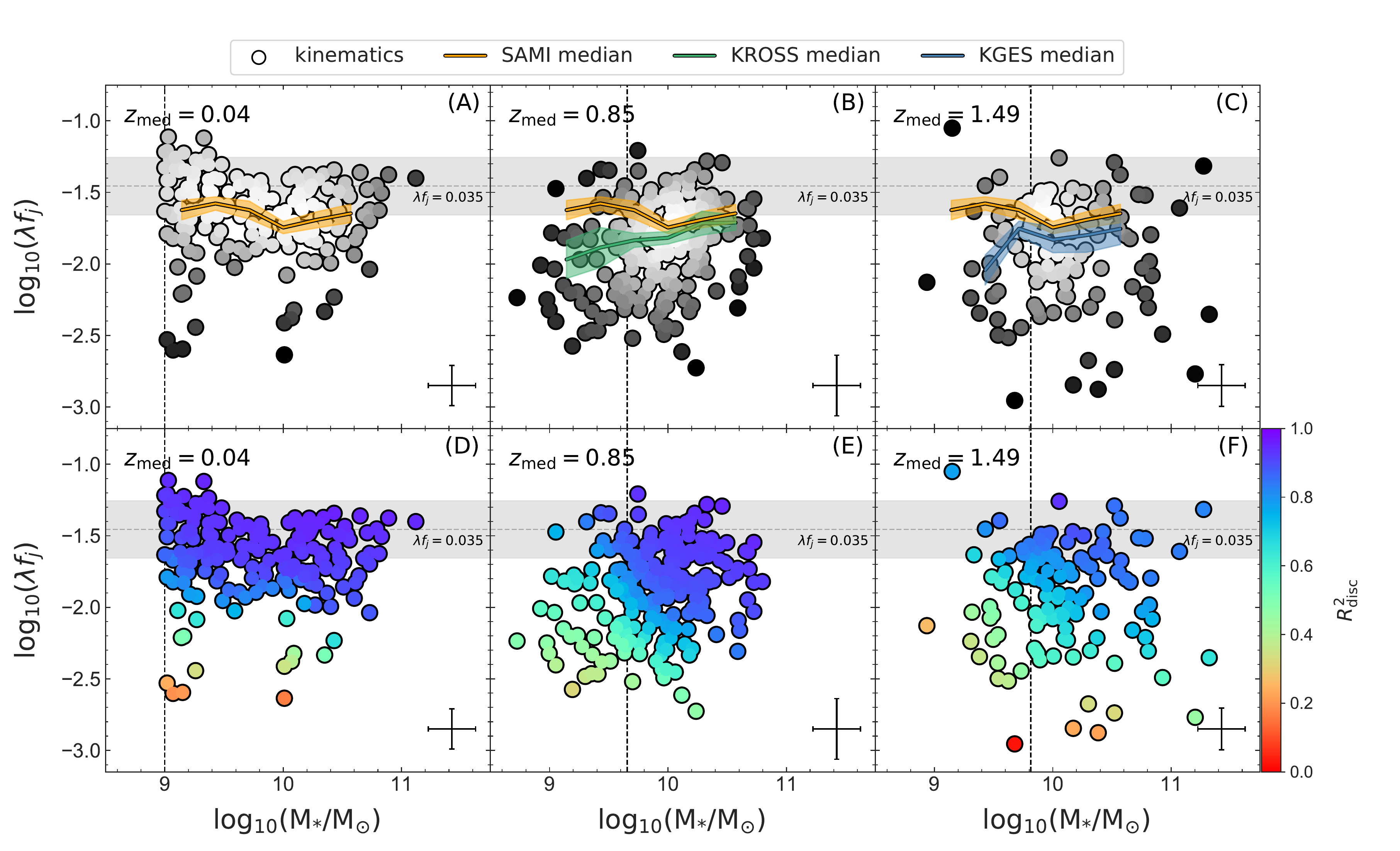}
\end{minipage}
\caption{%
The product of the halo spin ($\lambda$) and the ratio of the total specific angular momentum of the disc to that of the halo ($f_{j}$) as a function of stellar mass for {\it kinematics} sub-sample galaxies at $z \approx 0.04$, $\approx0.85$, and $\approx1.49$. In the top row the scatter points are colour-coded by the local spatial density of the points themselves (quantified via a Gaussian kernel density estimate and {\sc loess}-smoothed). In the bottom row the scatter points are colour-coded by the corresponding galaxy's ``disciness", $R^{2}_{\rm{disc}}$ (with {\sc loess} smoothing applied). The median trend for the SAMI galaxies, and its corresponding $1\sigma$ uncertainty, is displayed in each panel in the top row as the orange solid line and corresponding filled region, respectively. The median trends for the KROSS and KGES galaxies are displayed in respectively green and blue in the middle and right panel. The horizontal dashed grey line in each panel indicates the expected value if $f_{j} = 1$ and $\lambda = 0.035$ ($\pm 0.2$ dex, indicated via the horizontal filled grey region), the latter according to \citep{Maccio:2008}. In each panel we include a vertical dashed line representing the stellar mass above which each sample contain purely main sequence galaxies (see text for details). For main sequence star-forming galaxies, $\lambda f_{j}$ does not significantly depend on stellar mass or redshift. It does however strongly depend on how ``discy" a galaxy is (i.e. how disc-like its velocity field appears, as quantified by its $R^{2}_{\rm{disc}}$ value). %
     }%
\label{fig:spinfj_stellarmass}
\end{figure*}

\subsection{Linking star-forming galaxies to their haloes}
\label{subsec:galaxy_halo_link}

After examining the positions of our {\it kinematics} sub-sample galaxies in the $j_{*2.2_{\rm{C}}}$--$M_{*}$ plane, we now physically interpret our results in the context of galaxy formation theory. Our goal is to re-express the results presented in previous sections in terms of the fraction of the initial angular momentum retained by galaxies in our sample since their formation. This is a quantity that should be closely connected to their formation histories. We wish to determine (1) whether this quantity is a function of redshift for galaxies in our analysis, and (2) whether, at fixed redshift, the retention is dependent on total stellar mass. 

To do this we adopt the same simple model as applied to KROSS galaxies at $z\approx0.9$ by \citet{Harrison:2017}, linking the angular momentum of galaxies and their haloes, and based on analytical derivations given in \citet{Obreschkow:2014}. The model assumes that each galaxy, with specific angular momentum $j_{\rm{gal}}$, is embedded within a single spherical and isothermal cold dark matter halo that does not extend beyond its virial radius, and with specific angular momentum, $j_{\rm{halo}}$, and spin, $\lambda$. Rearranging Equation~7 of \citet{Harrison:2017} (which assumes a universal baryon fraction $f_{\rm{b}} = 0.17$), the product of the halo spin and the ratio of galaxy to halo angular momentum ($f_{j}=j_{\rm{gal}}/j_{\rm{halo}}$) can be expressed in terms of quantities that we may either estimate or evaluate directly for galaxies in our analysis, such that

\begin{equation}\label{eq:spinfj}
\lambda \times f_{j} = \frac{1}{2.95 \times 10^{4}}\cdot \left[\frac{H(z)}{H_{0}}\right]^{1/3} \cdot \frac{j_{\rm{gal}}/\rm{kpc\ km\ s}^{-1}}{(M_{*}/10^{11}\rm{M}_{\odot})^{2/3}}\cdot f_{s}^{2/3}
\end{equation}

\noindent where $H(z)$ and $H_{0}$ are respectively the Hubble Constant at redshifts $z$ and $z=0$, and $f_{s}$ is the ratio of stellar mass in the galaxy to the {\it initial} mass of gas in the halo ($M_{*}/M_{\rm{gas,0}}$). We proceed under the assumption that $j_{2.2_{\rm{C}}} \equiv j_{\rm{gal}}$ for galaxies in our {\it kinematics} sub-samples (and thus it is clear that the second term in Equation~\ref{eq:spinfj} is simply a renormalised version of the  $j_{*2.2_{\rm{C}}}$--$M_{*}$ normalisations, $j_{2.2_{\rm{C}}}/M_{*}^{2/3}$, discussed in previous sections). 

Whilst clearly not a direct measurement, the quantity $\lambda \times f_{j}$ can serve as a useful proxy for the fraction of angular momentum retained since a galaxy's formation. This is caveat to the assumption that $j_{\rm{halo}}$ and $\lambda$ have not changed since the halo's formation. It also requires us to know the value of the latter. In fact, as discussed later, we must actually {\it assume} a value for $\lambda$ for galaxies in our sample, in the absence of any direct measurements. For these reasons we stress that, for the results presented in this sub-section and Figure~\ref{fig:spinfj_stellarmass}, we are less concerned with the {\it absolute} value(s) of $\lambda \times f_{j}$ for galaxies in our analysis. Instead we focus on the {\it relative differences} between redshifts and stellar masses. These are hopefully less sensitive to the aforementioned assumptions. 

\subsubsection{Estimating the stellar mass-to-initial mass ratio}
\label{subsubsec:estimating_fs}

In the absence of a direct measurement of $f_{s}$ for our galaxies, we may instead calculate an approximation. Based on the stellar mass-halo mass relation measured by \citet{Dutton:2010b} for massive late-type galaxies at $z\approx0$, \citet{Harrison:2017} adopt a mass dependent analytical form for $f_{s}$, 

\begin{equation}
f_{s,\rm{D10}} = 0.29 \times \left(\frac{M_{*}}{5 \times 10^{10}\rm{M}_{\odot}}\right)^{1/2}\left(1 + \frac{M_{*}}{5 \times 10^{10}\rm{M}_{\odot}}\right).^{-1/2}
\end{equation}

\noindent However, this expression is based on measurements for massive galaxies only ($\log_{10}(M_{*}/\rm{M}_{\odot})\gtrsim9.5$--$10$), and the mass dependent aspect is primarily driven by those in the range $9.5 \lesssim \log_{10}(M_{*}/\rm{M}_{\odot}) \lesssim 10$; at higher stellar masses the ratio of stellar mass-to-halo mass for late-types measured by \citet{Dutton:2010b} is approximately flat (see Figure~1 of that work). 

Since our {\it kinematics} sub-samples extend to lower stellar masses (down to $\log_{10}(M_{*}/\rm{M}_{\odot}) \approx 9$), and we do not know a priori that there is indeed any stellar mass dependence of $f_{s}$ for our galaxies, we adopt a modified approach. We instead calculate $f_{s}$ using our total stellar mass measurements, as well as an estimate of the total gas mass ($M_{\rm{gas}}$) for each galaxy based on its star-formation rate and inverting the ``Kennicutt-Schmidt Law" \citep{Kennicutt:1998b}, that relates the gas surface density of a galaxy to its star-formation rate surface density (see Appendix~\ref{sec:gas_masses} for further details).

We thus estimate $f_{s}$ for our {\it kinematics} sub-sample galaxies in the following steps: 

\begin{itemize}
\item We first assume that $f_{s} \propto M_{*}/M_{\rm{b}}$, where $M_{\rm{b}} = M_{*} + M_{\rm{gas}}$ is the total baryonic mass. In other words, we assume $M_{\rm{b}} \propto M_{\rm{gas,0}}$, i.e. that the current baryonic mass is proportional to the initial gas mass within the halo when the galaxy first formed
\medskip
\item Next we calculate the normalisation itself, $X$ where $f_{s} = XM_{*}/M_{b}$, assuming that $X$ is the median value of the ratio $f_{s,\rm{D10}}/\left(\frac{M_{*}}{M_{\rm{b}}}\right)$ for massive ($\log_{10}M_{*}/\rm{M}_{\odot} > 10$), late-type ($n_{\rm{S}} < 1.5$) SAMI galaxies, i.e.\ for galaxies in our analysis with stellar masses, morphologies, and redshifts for which the \citet{Dutton:2010b} $f_{s,\rm{D10}}$ estimate is valid, and least mass dependent. 
\item Finally, we assume that the normalisation $X$ is valid for all galaxies in our {\it kinematics} sub-sample, i.e.\ that it does not change with stellar mass or redshift, and apply it to each to calculate $f_{s}$. 
\end{itemize}

\noindent In calculating $f_{s}$ for our galaxies in this way, we ensure that the normalisation is matched to the $f_{s,\rm{D10}}$ analytical estimate in the parameter space where this expression is valid, whilst simultaneously allowing for any deviation in the mass dependence of $f_{s}$ for our galaxies at lower stellar masses. This also implies that deviations between the two methods are on average relatively small (i.e., 0.06-0.15 dex) and mainly systematic, with $f_{s,\rm{D10}}$ 
slightly larger than $f_{s}$. The only exception is SAMI, where our technique produces values generally higher than $f_{s}$ for galaxies with stellar masses lower than  $\log_{10}M_{*}/\rm{M}_{\odot}=10$.

\subsubsection{Angular momentum retention of star-forming galaxies}
\label{subsubsec:angmom_retention}

In Figure~\ref{fig:spinfj_stellarmass} we plot our estimate of ($\log_{10}$) $\lambda \times f_{j}$ for our {\it kinematics} sub-sample galaxies at $z\approx1.5$, $\approx0.9$, and $\approx0.04$, as a function of their stellar mass. We colour code the scatter points on the upper and lower row by respectively their own surface density and the $R^{2}_{\rm{disc}}$ value for the corresponding galaxy. In each panel (each redshift) we indicate the nominal value $\lambda \times f_{j} = 0.035$. This is the value one would expect if the following were true: Firstly that the average halo spin $\langle \lambda \rangle = 0.035$ ($\pm 0.2$ dex), following the example of \citet{Romanowsky:2012}, \citet{Burkert:2016}, and \citet{Harrison:2017} and based on the average spin found by \citet{Maccio:2008} for halos spanning five orders of magnitude in mass in cosmological volume simulations with {\it WMAP}5 cosmologies. And secondly that $f_{j} = 1$, i.e.\ that the specific angular momentum of the galaxy is equal to that of its halo. 

With vertical dashed lines in Figure~\ref{fig:spinfj_stellarmass} we also indicate for each redshift the stellar mass above which our samples comprise only main sequence galaxies (i.e. the stellar mass above which the fraction of galaxies that are more than $5\sigma$ above the \citet{Schreiber:2015} main sequence, measured in running 0.2 dex bins of stellar mass, is consistently less than 15 per cent). All of the SAMI galaxies at $z\approx0.04$ considered in this work sit to the right of this line, by selection. This is true also for the vast majority of the $z\approx0.9$ and $z\approx1.5$ galaxies from KROSS and KGES, respectively. However, there is a small minority at each of these two redshifts that sit significantly above the main sequence. This is the result of the selection for H$\alpha$-detected sources and the H$\alpha$ flux detection limit that together effectively imposes a lower limit in star-formation rate, regardless of galaxies' stellar mass, meaning that at the very lowest stellar masses, galaxies have elevated {\it specific} star-formation rates and are more likely to reside above the main sequence of star formation.

It is important that we differentiate between those galaxies on and those above the main sequence in this way since, as discussed previously, in this work we are interested in ``typical" star-forming systems. Furthermore, there is evidence to suggest that galaxies above the main sequence of star-formation for their epoch may exhibit markedly different physical and kinematic properties to those that sit on it, including significantly reduced metallicities, enhanced gas fractions, more spatially concentrated star-formation, reduced stellar spin, and/or increased stellar bulge-to-total ratios \citep[e.g.][]{Magdis:2016,Morselli:2017,Elbaz:2018,Wang:2020}. On this basis, we focus our attention on those galaxies to the right of the vertical dashed lines in each panel of Figure~\ref{fig:spinfj_stellarmass}, where we may be sure we are considering purely main sequence systems.

The median $\log_{10}(\lambda \times f_{j})$ for main sequence galaxies (to the right of the vertical dashed line) is $-1.80 \pm 0.06$, $-1.79 \pm 0.03$, and $-1.63 \pm 0.02$ at $z\approx1.5$, $\approx0.9$, and $\approx0.04$, respectively. There is a respectively significant ($5.4\sigma$) and marginally significant ($2.8\sigma$) difference between the median $\log_{10}(\lambda \times f_{j})$ for $z\approx0.04$ main sequence galaxies and those at $z\approx0.9$ and $z\approx1.5$. However, these differences reduce if we further match the stellar mass range of the SAMI and KROSS galaxies at $z\approx0.04$ and $z\approx0.9$ to those of the KGES galaxies at $z\approx1.5$, i.e.\ if we consider {\it kinematics} sub-sample galaxies with $\log_{10}(M_{*}/\rm{M}_{\odot}) \geq 9.8$ (to the right of the dashed vertical lines in the rightmost column of Figure~\ref{fig:spinfj_stellarmass}) at each of the three redshifts. Then we find the median $\log_{10}(\lambda \times f_{j})$ at $z\approx0.9$ and $z\approx0.04$ to be respectively $-1.67 \pm 0.03$ and $-1.78 \pm 0.03$, and no significant difference between the median $\log_{10}(\lambda \times f_{j})$ across the three redshifts - only a marginal significant difference (of $2.6\sigma$) at most, between $z\approx0.9$ and $z\approx0.4$ galaxies. 

Considering the latter, stellar mass-matched median values, taken at face value, and assuming that $\lambda = 0.035$ for the halo of each galaxy and $f_{j} = 1$ at their initial formation, this would imply that, on average, main sequence, massive ($\log_{10}(M_{*}/\rm{M}_{\odot}) \geq 9.8$) star-forming galaxies at $z\approx1.5$, $\approx0.9$, and $\approx0.04$ have lost respectively $55^{+6}_{-7}$ , $53^{+3}_{-4}$, and $39^{+4}_{-5}$ per cent of their initial angular momentum over their lifetimes. We note that this is consistent with the findings of \citet{Harrison:2017}, who report a $\approx$40--$50$ per cent loss of initial angular momentum for star-forming ``discy" galaxies at $z\approx0.9$. 

Whilst qualitatively consistent with previous studies, we stress that the loss of initial angular momentum that we infer at each redshift only holds if the assumptions it is based on are strictly true. We therefore urge caution in directly interpreting deviation from $\lambda \times f_{j} = 0.035$ for galaxies in our sub-samples like this. 
The more important point is rather that, regardless of the absolute value of the median ($\log_{10}$) $\lambda \times f_{j}$ we measure at each redshift, the values do not significantly differ {\it between} redshifts -- perhaps implying that, on average, massive star-forming galaxies follow similar assembly pathways regardless of their cosmic epoch.

Furthermore, examining the distribution of scatter points at each redshift in Figure~\ref{fig:spinfj_stellarmass}, similar features to those discussed in \S~\ref{subsubsec:linear trends} in relation to Figure~\ref{fig:jstar_vs_Mstar} are apparent; the scatter at each redshift is skewed, with galaxies preferentially scattered towards lower $\log_{10}(\lambda \times f_{j})$ with respect to the main locus of scatter points. Similarly, as for Figure~\ref{fig:jstar_vs_Mstar}, we again see a stellar mass-dependent scatter for the KROSS galaxies at $z\approx0.9$. For these reasons we again rely on an examination of the running median trend at each redshift (as opposed to a straight line fit) in order to capture the average relationship between $\log_{10}(\lambda \times f_{j})$ and $\log_{10}(M_{*}/\rm{M}_{\odot})$ for massive, star-forming galaxies at each epoch. We find the median trends at each redshift to be consistent with one another to the right of the vertical dashed lines in each panel, after accounting for uncertainties. Thus, even after accounting for stellar mass, we still find no deviation in the median $\log_{10}(\lambda \times f_{j})$ for main sequence galaxies between redshifts. 

Similarly, the median trend at each redshift is approximately flat for purely main sequence galaxies (to the right of the dashed vertical line) at each redshift. We note that we do see an {\it apparent} trend in $\lambda \times f_{j}$ with stellar mass for the KROSS galaxies at $z\approx0.9$ when considered as a whole (i.e. galaxies both to the left and to the right of the vertical dashed line), however this is purely driven by the lowest stellar mass systems at that epoch, which themselves are significantly above the main sequence on average for their redshifts, as discussed. Thus we refrain from interpreting the positions of two distinctly different groups of galaxies (those above and on the main sequence) as a continual trend between $\lambda \times f_{j}$ and stellar mass. In fact, the distinct difference between the two groups is illuminating: the apparent stellar-mass dependence of $\lambda \times f_{j}$ at $z\approx0.9$ may in fact be purely the result of the finite H$\alpha$ flux limit for KROSS, which effectively acts to exclude main sequence galaxies at the very lowest stellar masses at that epoch. It may follow that these missing systems are likely to fall in the top left corner of the middle panels of Figure~\ref{fig:spinfj_stellarmass}. The positive trend with stellar mass in that case would then be simply an illusion due to selection effects at the lowest masses.

We also highlight the fact that, at fixed stellar mass, there is a strong dependence on $\log_{10}(\lambda \times f_{j})$ with $R^{2}_{\rm{disc}}$. And that, at fixed $R^{2}_{\rm{disc}}$, $\log_{10}(\lambda \times f_{j})$ is approximately flat with redshift. Furthermore, we  note that those galaxies (at $z\approx0.9$) significantly above the main sequence on average (i.e.\ points to the left of the vertical dashed line) also have low $R^{2}_{\rm{disc}}$, corresponding to their systematically lower $\lambda \times f_{j}$ in comparison to the remainder of the sample at that redshift. This confirms our conclusion from previous sections that the specific angular momentum of star-forming galaxies depends most strongly on their ``disciness", whilst being effectively independent of redshift. Specifically, it suggests that if massive star-forming galaxies retain some memory of, or link to, the angular momentum of their halos, they do so to the same extent at each redshift, and regardless of their stellar mass. Given also that the majority of our star-forming galaxies exhibit disc-like properties, regardless of redshift, this in turn may suggest that disc assembly may have followed a similar process throughout cosmic history. 

Finally, it is important to note that the results presented in this subsection are, in general, unaffected by our choice of approximation for $f_{s}$. Indeed, if we instead follow the method of \citet{Harrison:2017} and adopt $f_{s} = f_{s,\rm{D}10}$, we find no significant difference between the running median ($\log_{10}$) $\lambda \times f_{j}$ for main sequence galaxies (to the right of the dashed vertical lines in Figure~\ref{fig:spinfj_stellarmass}) at any of the three redshifts. In fact, the formal statistical significance of any differences decreases as the resultant $\lambda \times f_{j}$ are slightly elevated with respect to the those calculated using our preferred approximation for $f_{s}$. Similarly, for the same main sequence galaxies, adopting $f_{s} = f_{s,\rm{D}10}$ we again find no evidence for any significant mass dependence of $\lambda \times f_{j}$ at either $z\approx1.5$ or $z\approx0.9$. However, we do find a slight mass dependence for main sequence SAMI galaxies at $z\approx0.04$; if we adopt $f_{s} = f_{s,\rm{D}10}$, we measure a modest but significant slope of $0.27 \pm 0.04$ for the best fitting straight line to the positions of the SAMI galaxies (determined once again via {\sc hyperfit}) in the $\log_{10}(\lambda f_{j})$--$\log_{10}(M_{*}/\rm{M}_{\odot})$ plane at $z\approx0.04$.

\section{Conclusions}
\label{sec:conclusions}

We have presented the KMOS Galaxy Evolution Survey (KGES), a Durham University-led guaranteed time ESO KMOS study of the H$\alpha$ and [N {\sc ii}] emission from 288 $K$ band-selected galaxies at $1.2 \lesssim z \lesssim 1.8$. We characterised the properties of the KGES galaxies, and compared them to those of large samples of galaxies observed with IFS at $z\approx0.9$ by KROSS and $z\approx0.04$ by the SAMI Galaxy Survey. In this work:

\begin{itemize}
\item We confirmed that KGES galaxies represent typical star-forming galaxies for their epoch, residing on the main sequence of star-formation for their redshifts and stellar masses (Figure~\ref{fig:sfr_size_mass}), and with disc-like properties on average (\S~\ref{subsec:disciness}).
\item Combining the KGES galaxies with IFS samples of star-forming galaxies from KROSS and the SAMI Galaxy Survey, with exactly matched sample selections and analyses methods, and robustly accounting for differences in data quality between redshifts, we found that the fraction of discs (i.e.\ galaxies with both $R^{2}_{\rm{disc}} > 0.8$ and $v_{2.2_{\rm{C}}}/\sigma_{0_{\rm{C}}} > 3$) amongst the massive, star-forming population only modestly differs between $z\approx1.5$, $\approx0.9$, and $\approx0.04$ (by $\approx8$--$17$ percentage points across the {\it kinematics} sub-samples at each redshift, or $\approx3$--$31$ per cent within fixed bins of stellar mass across the three redshifts; Figure~\ref{fig:disc_fractions}). Instead it more strongly depends on stellar mass (differing by $\approx21$--$44$ per cent between the lowest and highest stellar mass galaxies in our sample at fixed redshift, depending on the redshift bin). 
\item We showed that the running median position of massive star-forming galaxies in the $j_{*2.2_{\rm{C}}}$--$M_{*}$ plane does not significantly differ between $z\approx1.5$, $\approx0.9$, and $\approx0.04$ (Figure~\ref{fig:jstar_vs_Mstar}).
\item Similarly, we showed that the median normalisation, calculated for individual galaxies as $j_{*2.2_{\rm{C}}}/M_{*}^{2/3}$, only varies slightly between the three redshifts - and only significantly so between $z\approx0.9$ and $z\approx0.04$ (differing by $0.13 \pm 0.03$ dex). Instead, we found that the median normalisation depended much more strongly on how disc-like a galaxy is, as judged by its $v_{2.2_{\rm{C}}}/\sigma_{0_{\rm{C}}}$ or $R^{2}_{\rm{disc}}$; the normalisation differed by $\approx0.4$--$0.6$ dex and $0.2$--$0.4$ dex between the lowest and highest bins of respectively $v_{2.2_{\rm{C}}}/\sigma_{0_{\rm{C}}}$ and $R^{2}_{\rm{disc}}$ at fixed redshift, depending on the redshift itself (Figure~\ref{fig:jstarMstar23_redshift}).
\item Lastly we interpreted our results in the context of a simple toy model, linking galaxies' specific angular momenta to that of their haloes. We found no strong evidence to suggest that the product of the halo spin and the ratio of the galaxy's specific angular momentum to that of its halo, $\lambda \times f_{j}$, is dependent on redshift or stellar mass for massive, star-forming galaxies on the main sequence at $z\approx1.5$, $z\approx0.9$, and $z\approx0.04$ (Figure~\ref{fig:spinfj_stellarmass}). We found instead that it depends most strongly on how disc-like a galaxy is, regardless of mass or redshift.
\end{itemize}

Our results suggest that the inferred link between the angular momentum of galaxies and their haloes does not depend on stellar mass or redshift for star-forming galaxies. Combined with the fact that we find, at-most, only modest differences in the disc fraction of the star-forming galaxy population between redshifts, this in turn suggests that massive star-forming galaxies may have followed similar assembly pathways over the past $\approx10$ Gyr.

\section*{Acknowledgments}
We thank the referee for useful comments which improved the quality of this manuscript. 
ALT acknowledges support from a Forrest Research Foundation Fellowship, Science and Technology Facilities Council (STFC) grants ST/L00075X/1, and the ERC Advanced Grant DUSTYGAL (321334). ALT and AP acknowledge support from STFC (ST/P000541/1). SG acknowledges the support of the Science and Technology Facilities Council through grant ST/N50404X/1 for support and the Cosmic Dawn Center of Excellence funded by the Danish National Research Foundation under then grant No. 140.  LC is the recipient of an Australian Research Council Future Fellowship (FT180100066) funded by the Australian Government. AMS, IS, RMS and AP acknowledge support from STFC (ST/T000244/1). UD acknowledges the support of STFC studentship (ST/R504725/1). Parts of this research were supported by the Australian Research Council Centre of Excellence for All Sky Astrophysics in 3 Dimensions (ASTRO 3D), through project number CE170100013.
\section*{Data Availability}
The data underlying this article are available in the ESO archives for KGES and KROSS and at \url{https://docs.datacentral.org.au/sami/} for the SAMI survey. 
Derived properties for the KGES galaxies are available in the article and in its online supplementary material. 
Derived properties for the SAMI and KROSS samples have been presented in \cite{Tiley:2018}.



\bibliographystyle{mnras}
\bibliography{paper.bib} 



\newpage

\appendix
\section{Table of Values}
\label{sec:appendixtable}

In Table~\ref{tab:kgesvals} we tabulate the key properties of the KGES galaxies. A machine-readable version of this table will be made publicly available online in full, upon publication.

\begin{sidewaystable}
\vspace{8cm}
\centering
\caption{\normalsize Key properties of the KGES sample galaxies discussed in this work.}\label{tab:kgesvals}
\small
\begin{tabular}{ lllllllccccccc}
\hline
KGES  & R.A. & Dec & Redshift & H$\alpha$- & H$\alpha$- & AGN & {\it kinematics} & $M_{*}$ & $R_{50}$ & $\rm{SFR}_{\rm{H}\alpha}$ & $v_{2.2_{\rm{C}}}$ & $\sigma_{0_{\rm{C}}}$ & $j_{2.2_{\rm{C}}}$ \\
SURVEY ID  & (deg) & (deg) &  & detected  & resolved &  &  & (10$^{10}$ $\rm{M}_{\odot}$) & (kpc) & ($\rm{M}_{\odot} \rm{yr}^{-1}$) & (km s$^{-1}$) & (km s$^{-1}$) & (kpc km s$^{-1}$)\\
(1) & (2)  & (3) & (4) & (5) & (6) & (7) & (8) & (9) & (10) & (11) & (12) & (13) & (14) \\
\hline
KGES\_1 & 53.134483 & -27.770931 & 1.552 & True & False & False & False & 3.63 & 2.76 & 2.69 & - & - & - \\
KGES\_2 & 53.065608 & -27.767825 & 1.539 & True & True & False & True & 6.55 & 4.54 & 59.4 & 148.8  & 69.6 & 790.5 \\
KGES\_3 & 53.110238 & -27.763039 & 1.470 & True & True & True & False & 6.41 & 6.06 & 12.9 & 183.3  & 43.1 & 1312.8 \\
. & . & . & . & . & . & . & . & . & . & . & .  & . & . \\
. & . & . & . & . & . & . & . & . & . & . & .  & . & . \\
. & . & . & . & . & . & . & . & . & . & . & .  & . & . \\
\hline
\multicolumn{14}{l}{%
\begin{minipage}{24cm}%
\medskip
\normalsize Notes. {\bf (1)} KGES survey ID.  {\bf (2)} Right ascension. {\bf (3)} Declination. {\bf (4)} Redshift (spectroscopic if detected in H$\alpha$, photometric if not). {\bf (5)} H$\alpha$-detected flag. {\bf (6)} H$\alpha$-resolved flag. {\bf (7)} Candidate AGN host flag. {\bf (8)} {\it kinematics} sub-sample membership flag. {\bf (9)} Stellar mass. {\bf (10)} Stellar light effective radius. {\bf (11)} H$\alpha$-derived total (i.e.\ aperture- and attenuation-corrected) star-formation rate. {\bf (12)} Inclination- and beam smearing-corrected rotation velocity at $1.31R_{50}$. {\bf (13)} Beam smearing-corrected velocity dispersion. {\bf (14)} Total stellar specific angular momentum. %
\end{minipage}%
}\\
\end{tabular}
\end{sidewaystable}

\newpage
\section{Estimating total gas masses}
\label{sec:gas_masses}

For each galaxy we estimate its gas mass within $R_{50}$ by inverting the ``Kennicutt-Schmidt Law" \citep{Kennicutt:1998b}, that relates the gas surface density of a galaxy to its star-formation rate surface density such that 

\begin{equation}
\frac{\Sigma_{\rm{gas},50}}{\rm{M}_{\odot}\ \rm{pc}^{-2}} = \left(\frac{4 \times 10^{4}\ \Sigma_{\rm{SFR},50}}{\rm{M}_{\odot}\ \rm{yr}^{-1}\ \rm{kpc}^{-2}}\right)^{0.714}\,\,,
\label{eq:gasdensity}
\end{equation}

\noindent where $\Sigma_{\rm{gas},50}$ and $\Sigma_{\rm{SFR},50}$ are respectively the gas surface density and star-formation rate surface density within a circular aperture with radius $R_{50}$. We calculate the latter as 

\begin{equation}
\frac{\Sigma_{\rm{SFR},50}}{\rm{M}_{\odot}\ \rm{yr}^{-1}\ \rm{kpc}^{-2}} = \frac{\rm{SFR}}{\rm{M}_{\odot}\ \rm{yr}^{-1}} \cdot \frac{\rm{kpc}^{2}}{2 \pi \rm{C}_{\rm{IMF}} R_{50}^{2}}\,\,,
\label{eq:sfrdensity}
\end{equation}

\noindent where for the KROSS and KGES galaxies $\rm{SFR}$ is the total, H$\alpha$-derived ($\rm{SFR}_{\rm{H}\alpha}$), as calculated calculated in \S~\ref{subsec:totfluxes_and_sfrs}, and for SAMI galaxies it is the {\sc magphys} derived quantity measured by \citet{Davies:2016}.

We convert $\Sigma_{\rm{gas},50}$ to a {\it total} gas mass ($\rm{M}_{\rm{gas}}$) as

\begin{equation}
\frac{\rm{M}_{\rm{gas}}}{\rm{M}_{\odot}} = \frac{2\ \Sigma_{\rm{gas},50}}{\rm{M}_{\odot}\ \rm{pc}^{-2}} \cdot \frac{\pi\ R_{50}^{2}}{\rm{pc}^{2}}\,\,.
\label{eq:gasmass}
\end{equation}

\noindent In other words, the total gas mass is calculated as twice the mass of gas within $R_{50}$ that is inferred from each galaxy's H$\alpha$-derived SFR.

\bsp	
\label{lastpage}
\end{document}